\documentclass[11pt,a4paper]{article}
\pdfoutput=1
\usepackage{jheppub}  
\usepackage{amsfonts,amsbsy,amsmath,amssymb}
\usepackage{color}
\usepackage{graphicx}
\usepackage[colorinlistoftodos]{todonotes}
\usepackage{booktabs}
\usepackage{caption}
\usepackage{subcaption} 
\usepackage{multirow} 
\usepackage{algorithm}
\usepackage[noend]{algpseudocode}
\usepackage[section]{placeins}
\usepackage[normalem]{ulem}

\newcommand{\comment}[1]{}
\newcommand{\thetat}{\tilde{\theta}}
\newcommand{\kbar}{\bar{k}}
\newcommand{\Tr}{\text{Tr}}
\newcommand{\tmin}{t_{\mathrm{min}}}

\newcommand{\dof}{\mathrm{dof}}
\newcommand{\HG}{\hat{\Gamma}}
\newcommand{\be}{\begin{equation}}
\newcommand{\ee}{\end{equation}}
\newcommand{\ba}{\begin{array}}
\newcommand{\ea}{\end{array}}
\newcommand{\baa}{\begin{array}}
\newcommand{\eaa}{\end{array}}
\newcommand{\bea}{\begin{eqnarray}}
\newcommand{\eea}{\end{eqnarray}}

\newcommand{\kb}{\bar{k}} 

\newcommand{\vpt}{\vec{p}}

\newcommand{\lef}{\tilde l}

\newcommand{\ttheta}{\tilde \theta}

\newcommand{\Ok}{\Omega_{[\vec n]}}

\newcommand{\vn}{\vec n}
\newcommand{\cE}{{\cal E}}
\newcommand{\pz}{p_0}
\newcommand{\zmin}{Z_{\mathrm{min}}}
\newcommand{\amax}{A_{\mathrm{max}}}

\title{The spectrum of 2+1 dimensional Yang-Mills theory on a twisted spatial torus}

\author[a]{Margarita Garc\'{i}a P\'erez,}
\author[a,b]{Antonio Gonz\'alez-Arroyo,}
\author[c]{Mateusz Koren,}
\author[d,e]{\\ Masanori Okawa}

\affiliation[a]{Instituto de F\'{i}sica Te\'orica UAM-CSIC, Nicol\'as
  Cabrera 13-15, Universidad Aut\'onoma de Madrid, E-28049 Madrid, Spain}
\affiliation[b]{Departamento de F\'{i}sica Te\'orica, 
M\'odulo 15,  Universidad Aut\'onoma de Madrid, Cantoblanco, E-28049 Madrid, Spain}
\affiliation[c]{John von Neumann Institute for Computing (NIC), DESY, 
Platanenallee 6, D-15738 Zeuthen, Germany}
\affiliation[d]{Graduate School of Science, Hiroshima University,
  Higashi-Hiroshima, Hiroshima 739-8526, Japan}
\affiliation[e]{Core of Research for the Energetic Universe, Hiroshima University,
                     Higashi-Hiroshima, Hiroshima 739-8526, Japan}

\emailAdd{margarita.garcia@uam.es}
\emailAdd{antonio.gonzalez-arroyo@uam.es}
\emailAdd{mateusz.koren@desy.de}
\emailAdd{okawa@hiroshima-u.ac.jp}

\abstract{ We compute and analyse the low-lying spectrum of 2+1 dimensional 
$SU(N)$ Yang-Mills theory on a spatial torus of size $l\times l$ with twisted
boundary conditions. This paper extends our previous work~\cite{Perez:2013dra}. 
In that paper we studied the sector with non-vanishing electric flux and
concluded that the energies only depend on the parameters through two 
combinations: $x=\lambda N l /(4\pi)$ (with $\lambda$  the 't Hooft coupling)
and the twist angle $\thetat$ defined in terms of the magnetic flux piercing 
the two-dimensional box. Here we made a more complete study and we are able 
to condense our results, obtained by non-perturbative lattice methods, into a
simple expression which has important  implications for the absence of 
tachyonic instabilities, volume independence and non-commutative field theory.
Then we extend our study to the sector of vanishing electric flux. We conclude
that the onset of the would-be large-volume glueball states occurs at an 
approximately  fixed value of $x$, much before the stringy torelon states have
become very massive. 
}


\preprint{%
{\flushright
IFT-UAM/CSIC-18-71\\
FTUAM-18-19\\
DESY-18-108\\
HUPD-1804 \\
}}

\begin{document}


\maketitle

\section{Introduction}
The present paper extends our previous study ~\cite{Perez:2013dra,Perez:2014sqa,Perez:2014jra} on the behaviour of pure SU(N) Yang-Mills theory in 2+1 dimensions, where  space is compactified as a 2-dimensional torus with 't Hooft twisted boundary conditions. We focus upon the dependence of the spectrum of the theory on the parameters that define it, namely the torus size $l$ (restricted to be rectangular of  equal length in both directions), the number of colours $N$,  and the magnetic flux $k$ (a modulo $N$ integer, coprime to $N$) introduced by the boundary conditions. 
There are certain results and observations that suggest that these parameters conspire to produce a simpler result. The dependence on the value of the coupling constant $\lambda$ ('t Hooft coupling) is implicitly contained, since it is used as the unit of energy or inverse length. In our first paper on the subject~\cite{Perez:2013dra}, we studied the spectrum for small torus sizes using perturbation theory. Our study showed that the individual energies depend only on two combinations of parameters: the  product of $N$ times the torus lateral size and the following angular variable: $\ttheta =2 \pi\bar{k}/N$, where $\bar{k}$ is the modular multiplicative inverse of $k$ ($k\bar{k}=1 \bmod N$). The second parameter  $\ttheta/(2\pi)$ is a rational number, but continuity of the energies with respect to it takes place in perturbation theory. Our analysis then used the lattice regularized version of the model to study how the energies evolve with  the torus size from the perturbative results to the expected confinement behaviour. The results  supported that the same two combinations described the dependence of the energies for all torus sizes and that continuity in $\thetat$ still applied.  

The dependence on the product  $lN$, or rather the dimensionless ratio $x=\lambda lN/(4 \pi)$, is connected to the phenomenon of volume independence at large $N$, since in that limit $x$ goes to infinity for any value of $l$.  
The question has been studied in 4 dimensions going back to the old observation of Eguchi and Kawai~\cite{Eguchi:1982nm} that Schwinger-Dyson equations are independent of the volume at large $N$. However, the proof assumes that center symmetry remains unbroken, a feature that turns out to be wrong  in the weak coupling regime~\cite{Bhanot:1982sh}. 
Several ideas have been presented over the years to solve this problem. A simple proposal, which is adopted in this paper, is to use 't Hooft twisted boundary conditions~\cite{GonzalezArroyo:1982ub,GonzalezArroyo:1982hz}. Other alternatives have been proposed, such as that of adding fermions in the   
adjoint representation~\cite{Kovtun:2007py,Basar:2013sza} and other modifications of the reduction idea~\cite{Unsal:2008ch}. Recent tests give compelling evidence that indeed these methods do rescue the idea of volume independence at least in 4 dimensions~\cite{Gonzalez-Arroyo:2014dua,GonzalezArroyo:2012fx}. The old studies only required that the flux $k$ associated to the twisted boundary conditions should be irreducible. However, more recent studies~\cite{Ishikawa,Bietenholz:2006cz,Teper:2006sp,Azeyanagi:2007su} showed problems and signs of symmetry breaking, indicating that  the choice of the magnetic flux $k$ is crucial to avoid instabilities and to reduce finite $N$ corrections. In Ref.~\cite{GonzalezArroyo:2010ss} two of the present authors proposed  that $k/N$ and $\bar{k}/N$ have to be kept finite (and beyond a certain threshold) to preserve the validity of volume independence at large $N$. Although there is a rationale behind the bound on $k/N$, there is no rigorous argument and  no precise estimate of its value. However, the practical attitude has been to show that  volume independence can be used to obtain precise results about the large $N$ infinite volume theory. This is similar to the use of perturbation theory in theories where no rigorous proof of summability is available.  Our study in the simpler 2+1 dimensional case is aimed precisely at clarifying some of these points. Not much work has been done in this system in this context apart from that of Narayanan and Neuberger~\cite{Narayanan:2003fc,Narayanan:2007ug}. 

There is another perspective from which our work has a long-standing interest: non-commutative field theories~\cite{Douglas:2001ba}. Indeed, twisted boundary conditions and non-commutative field theories are intimately connected. The first appearance in the literature of the action and Feynman rules of gauge and non-gauge non-commutative field theories was as a generalization of the volume reduced twisted models~\cite{GonzalezArroyo:1983ac}. The twisted Eguchi-Kawai model was also used as lattice-regularized version of non-commutative Yang-Mills theory~\cite{Ambjorn:2000cs,Ambjorn:2000nb}. As a matter of fact, Morita equivalence implies that gauge theories on the torus with twisted boundary conditions are particular cases of non-commutative field theories on the non-commutative torus with special values of the non-commutativity parameter. Indeed, one in which a dimensionless combination of the torus size and the non-commutativity parameter takes rational values.  
Our observations of Ref.~\cite{Perez:2013dra,Perez:2014sqa} appear natural within this context, since $lN$ is the size of the non-commutative torus and $\ttheta/(2\pi)$ the rational dimensionless non-commutativity parameter. Several studies showed that although the torus size tames down the IR/UV mixing~\cite{Minwalla:1999px}, new instabilities, called {\em tachyonic instabilities}, could appear~\cite{Hayakawa:1999zf,Guralnik:2002ru}.  These are  of the type associated with spontaneous breaking of centre symmetry and breakdown of volume independence. Our perturbative analysis of Ref.~\cite{Perez:2013dra,Perez:2014sqa}  agrees with this conclusion, but also shows that a suitable choice of the flux $k$ can help to avoid them. Furthermore, our work  also serves to extend this analysis to larger torus sizes for which the perturbative calculation breaks down. Hence, our analysis is capable of addressing some of the questions raised long time ago within non-commutative field theories beyond the domain of perturbation theory. Here we signal out very specially the ideas and  hypothesis  formulated in Ref.~\cite{Guralnik:2001pv,AlvarezGaume:2001tv}.

Having set the general context  of our present study, we go into an overview of the specific goals of the present paper. The spectrum of states can be split into sectors corresponding to different values of 't Hooft 
electric flux, a two-dimensional vector of integers modulo $N$. Our previous papers only studied certain sectors with non-vanishing values of this electric flux. In this paper we will extend the range of our study to cover all electric flux sectors for a wider interval of values of $N$, torus size and flux parameters. This will allow us to achieve a semi-quantitative phenomenological description of the dependence of these energies on the parameters. In addition, our study will also include the sector with vanishing electric flux, {\em the glueballs}.  
As we will see, the possibility of obtaining large-volume results from our large-$N$ analysis, as implied by the volume independence hypothesis, seems challenging. The reason being the presence of an ever increasing number of torelon pair states whose mass only grows with $l$, not $lN$. Our results provide a consistent solution to this puzzle.  

The layout of the paper is as follows. In Section~\ref{generalities} we present the necessary background to the problem, including previous results and a summary of what is known and expected for the system. In the next two sections  we present our main results, dealing with  non-zero (sect.~\ref{s.torelon}) and zero electric flux states (sect.~\ref{s.glueball}) respectively. To facilitate reading we have divided each section into three main parts: a preamble, a part in which the non-perturbative results are presented, and another one where the conclusions from these results are extracted. Within the same spirit, we have concentrated all technical aspects about the methodology used, which is by itself quite challenging, to an appendix. A list of tables provide the actual measured numbers for the sake of other researchers who might want to analyse the huge wealth of information that we have obtained. The paper closes with a brief conclusion section which gives a big overview and lists open problems and paths for improvement.

\section{$SU(N)$ gauge theory on a spatial two-dimensional twisted box}
\label{generalities}
We will be considering a $SU(N)$ gauge theory defined on $T^2 \times R$. For simplicity, the spatial 2-dimensional torus is 
considered to be symmetric and of period $l$.  The gauge potential satisfies twisted boundary conditions, given by
\be
A_\mu(x + l \hat e_i) = \Gamma_i A_\mu(x)  \Gamma_i^\dagger,
\ee
with $SU(N)$ matrices $\Gamma_i$ subject to the consistency condition:
\be
\Gamma_1 \Gamma_2 = e^{i \frac{2 \pi  k}{N}} \Gamma_2 \Gamma_1,
\label{eq.consist}
\ee
derived from imposing univaluedness of the gauge potential under displacement along the two cycles of the torus.
We will be considering the case of the so-called irreducible twists, in which the magnetic flux integer $k$ is taken to be coprime with $N$ \cite{ga:torus}. 
In that case, Eq.~\eqref{eq.consist} defines the matrices $\Gamma_i$ uniquely modulo global gauge transformations. 

Twisted boundary condition on a torus were introduced by 't Hooft \cite{'tHooft:1979uj,'tHooft:1980dx} as a way to induce topological (chromo-) electric and
magnetic fluxes in Yang-Mills theories. This is best understood in the Hamiltonian formalism in the $A_0=0$ gauge. 
In this set-up, time-independent large gauge transformations with non-trivial periodicity:
\be
\Omega_{[\vec n]}(\vec x + l \hat e_i) =  e^{i \frac{2 \pi n_i}{N}}\Gamma_i \Omega_{[\vec n]} (\vec x) \Gamma_i^\dagger,
\ee
act as symmetries of the Hamiltonian and allow to classify the states in the Hilbert space according to the transformation properties
under the $\Ok$:
\be
U(\Ok) |\psi_{\vec e} \rangle = e^{i \frac{2 \pi  \vec n \cdot \vec e}{N}} |\psi_{\vec e} \rangle.
\ee
States are thus classified by a 2-dimensional vector of integers $\vec e$ defined modulo $N$: the electric flux vector. 
In this way, the Hilbert space decomposes into $Z_{N}^2$ disjoint sectors parameterized by the value of the electric flux. 
The vacuum and the glueballs live in the sector with zero electric flux, while Polyakov loop operators with non trivial winding acting on the vacuum generate
non-zero electric flux states, the torelons~\cite{Michael:1986cj}.  
As mentioned in the introduction, the purpose of this paper is to present the results of a non-perturbative analysis of the volume and $N$ dependence of the spectrum 
in the different sectors, extending to the zero electric flux sector the results obtained in Ref.~\cite{Perez:2013dra} for non-zero fluxes.  
This will be done in sections \ref{s.torelon} and \ref{s.glueball} where we will analyse the spectrum obtained from a lattice Monte-Carlo simulation of the 2+1 dimensional system.
Before doing that, it is instructive to discuss what is the expected volume dependence based on what we know from perturbation theory and from 
the large volume, confinement regime.  

Asymptotic freedom implies that perturbation theory is a good approximation for small torus sizes. The calculations with twisted boundary conditions are easily performed when using an appropriate basis of the   $SU(N)$ Lie algebra~\cite{GonzalezArroyo:1982hz}. In this basis the vector potential can be expanded in a modified Fourier expansion:
\be
A_i(x) = \frac{1}{ l} \sum_{\vec p}^\prime e^{i \vec p \cdot \vec x} \hat A_i(t,\vec p\,) \,\HG(\vec p\,),
\label{eq.fourier}
\ee
where the momentum dependent matrices $\HG(\vec p\,)$ satisfy
\be
\Gamma_i\HG(\vec p\,)\Gamma_i^\dagger = e^{i l p_i} \HG(\vec p\,).
\ee
Using this formula it is easy to verify that the twisted boundary conditions amount to the quantization of momenta 
\be
\vec p =   ( n_1 , n_2 )\,  \pz, \quad n_i \in \mathbb{Z},
\label{eq.momq}
\ee
where the quantum of momentum $\pz=2 \pi /(l N)$.
This corresponds to the standard minimum momentum for an effective box size $\lef\equiv lN$.

The $\HG(\vec p\,)$ can be written explicitly 
as follows:
\be
\HG(\pz \vec n ) = \frac{1}{\sqrt{2N}} \, e^{i \alpha(\vpt)} \, \Gamma_1^{-\bar k n_2}  \Gamma_2^{\bar k n_1},
\label{eq.hg}
\ee 
where $\kbar$ is the modular multiplicative inverse of $k$:
\be
k\kbar = 1\;(\text{mod } N).
\label{eq.kb}
\ee
For irreducible twists, there are $N^2$ independent $\HG$  matrices of this sort that can be chosen as those with $n_i$ 
taking values from 0 to $N-1$. The one corresponding 
to $\vec n = \vec 0$ is proportional to the identity, while the remaining $N^2-1$ matrices are traceless and provide a basis for the (complexified) $SU(N)$ Lie algebra. The Fourier coefficients $\hat A_i(t,\vec p\,)$ are complex and satisfy a hermiticity condition similar to the standard one, which restricts the vector potentials to live in the standard real $SU(N)$ Lie algebra. 
The primed momentum sum in Eq.~\eqref{eq.fourier} runs over all momenta 
that lead to traceless $\HG$  matrices, excluding those with non-zero trace corresponding to $\vec n = \vec 0$ (mod $N$). Notice that this restriction implies in particular that zero-momentum is forbidden in the twisted box, and the minimum value of $|\vec p|=\pz$.

Equipped with the previous formalism it is very easy to compute the spectrum to leading order of perturbation theory. At this order the system can be described as a gas of free massless gluons whose energy is just given by the modulus of its momentum $E=|\vec p|$. The ground state or vacuum is the state with no gluons and has zero energy.
As a result of what we discussed in the previous paragraph the spectrum has a gap corresponding to a single gluon of minimum momentum  $\pz$ (it is 4-fold degenerate). If we write the energy in units of 't Hooft  dimensionful coupling $\lambda$  this  gap becomes $1/2x$, where 
\be
x= \frac {\lambda N l} {4 \pi}
\ee
is the quantity introduced in  Ref.~\cite{Perez:2013dra}, and which we argued is the relevant dynamical variable, setting the scale  both in the perturbative and in the non-perturbative regimes. We will review our arguments below. 

But, how does this leading-order spectrum correspond with the electric flux sectors mentioned earlier? To see this let us consider a spatial Polyakov loop  with non-trivial winding in the spatial torus. With twisted boundary conditions the Polyakov loop adopts the form:
\be
{\cal P}(\gamma)\equiv {\rm Tr} \Big( T\exp \Big\{ -i \int_\gamma dx_i A_i(x)\Big \}\, 
\Gamma_2^{\omega_2} \Gamma_1^{\omega_1}\Big ),
\ee
where $\gamma$ is a closed curve on the 2-torus and $\vec{\omega}$ is the
corresponding winding number. 
The state obtained  by acting with this operator over the vacuum has an electric flux given by $\vec{\omega}$ modulo $N$. Expanding the ordered exponential of the Polyakov loop and using the Fourier expansion of the vector potential given earlier,  we conclude that a gluon of momentum $\vec p=  \vec n \pz$  carries electric flux given by
\be
\vec{e} =  (n_2,- n_1) \kbar \quad \bmod{N}.
\label{eq.edef}
\ee
The fact that  $\kbar$ is defined modulo $N$ can be expressed by saying that the quantity 
 $\thetat$ defined as:
\be
\thetat = \frac{2\pi \kbar}{N}\, 
\label{eq.thetat}
\ee
is an angle. This is the other main quantity introduced in Ref.~\cite{Perez:2013dra} to describe the spectra.

At this stage it must be said that our two main quantities have a natural interpretation within the non-commutative field theory description. The 
effective size $\lef$ is just the size parameter of the non-commutative torus and $\thetat/(2 \pi)=\kbar/N$ is the dimensionless non-commutativity parameter.

Inverting the previous formula Eq.~\eqref{eq.edef} one can determine the minimum momentum corresponding to each electric flux $\vec e$ as follows $\vec p_c(\vec e)= (\pm \bar n_1,\pm \bar n_2)\, \pz $ with 
\be
\bar n_i = N \, \Big|\Big|\epsilon_{ij}\, \frac{k \,  e_j} {N}\Big|\Big| \, ,
\label{eq.ndef}
\ee
where we have followed the notation in \cite{Chamizo:2016msz}, with $||s||$ denoting the distance of the real number $s$ to the nearest integer. 
Hence, the lowest perturbative energy, corresponding to a one gluon state carrying momenta $\vec p=  \vec n \pz$ with  $|\vec n|=1$, belongs to the electric flux sectors    $\vec e = (\pm\kb, 0)$ and $(0, \pm\kb)$ -- see Eq.~\eqref{eq.edef}.
For multiple gluon states the momenta add up and hence, so does the corresponding electric flux. Notice that conservation of electric flux directly follows from conservation of momenta. It can happen that there are multigluon states degenerate with the single gluon state. For example, the state with two gluons of momenta $ (\pz,0)$ each, is degenerate with the state of one gluon of momenta $(2 \pz,0)$. This degeneracy only occurs for collinear gluons. 

Excited states in the zero electric flux sector can be obtained to leading order in perturbation theory as multigluon states. The first excited  state (the mass gap) is given by a pair of gluons of opposite momenta equal to the minimum one $|\vec p|=\pz$. In $\lambda$ units the energy is just $1/x$. Generically we might call these states glueballs, since they have the same quantum numbers as the corresponding states at large volumes. However, later we will reserve that name to the states that are present in the infinite volume theory, while these other will be referred as torelon pairs.
In the following sections we will present the results of our study of the spectrum separating the cases of non-vanishing and vanishing electric flux.

\section{The torelon spectrum (non-zero electric flux)}
\label{s.torelon}

\subsection{General considerations}
In this section we will deal with the case of non-vanishing electric flux. The corresponding energy eigenstates were called torelons in Ref.~\cite{Michael:1986cj}. This was already studied in Ref.~\cite{Perez:2013dra}, but we will present additional results which will allow us to draw certain conclusions from them.

We will label the torelon energies in units  of the 
't Hooft coupling $\lambda$ for each momentum value $\vec p=\vec p_c(\vec e) =\vec n \pz$   by the symbol ${\cal E}_{ \vn}$. 
In Ref.~\cite{Perez:2013dra} we computed the next to leading order perturbative contribution  to these energies coming from self-energy gluon diagrams. This combines with the leading order result into the following expression:
\be
 {\cal E}^2_{\vn}\, (x,\thetat) =    \frac{ |\vn|^2 }{4 x^2}  -   G \Big(\frac{\thetat \vec n}{2 \pi} \Big) \, \frac{1}{ x }
\label{efluxenpert}
\ee
with the part in $1/x$ representing the gluon self-energy  given in terms of the function:
\be
G ({\vec z} ) = -\frac{1}{ 16 \pi^2} \int_0^\infty
\frac{dt}{  \sqrt{ t}} \,
\Big(\theta_3^2(0,it) -\theta_3(z_1,i t) \, \theta_3(z_2,i t)- \frac{1}{  t}\Big)\quad ,
\label{eq:Gtheta}
\ee
with $\theta_3$ the Jacobi theta function~\cite{Tata}:
\be
\theta_3(z,it) = \sum_{k \in {\mathbf Z}} \exp\{-t \pi k^2 + 2 \pi i k z\}\quad .
\ee
One can see that for each value of $\vec n$ the energy square depends only on the two combinations of the arguments $x$ and $\thetat$. Notice that the argument of the function $G$ is just the (rotated) electric flux over $N$:  $ \epsilon_{ij} \thetat n_j/(2 \pi)= e_i /N$.
 
There are some interesting observations following from Eq.~\eqref{efluxenpert}. The first is that the simple form appears when writing the energy square and not the energy. Since the first term is the momentum of the gluon square, the second can be interpreted as the mass square. However, the latter is actually negative and would eventually drive negative the energy square at some finite value of $x$. This would signal a phase transition, that for obvious reasons is called a tachyonic instability. When the  problem was studied many years ago within the context of non-commutative field theories~\cite{Hayakawa:1999zf,Guralnik:2002ru},
it was correctly understood that the new phase would be one in which there is condensation of electric flux sectors  into the vacuum and  centre symmetry spontaneously broken. 

The arguments in favour of a tachyonic instability are non-rigorous because they are based on a truncated perturbative expansion. However, the function $G(\vec z)$ has a pole when the argument takes integer values, and hence the self-energies for large $N$ and small electric fluxes grow, making the transition point occur at small values of $x$ when the effect of higher order terms might be considered
negligible. Hence, one of the goals of  our study is that of addressing  whether such a phase transition exists or not. The result has implications for the non-commutative field theory program. 

With or without transition, when the torus size  is large the system should return to the confinement phase in which these electric flux energies would grow linearly with the torus period and eventually decouple from the rest of the system. The behaviour is expected to be that following from an effective string description. The two leading terms at large $x$ for the energy square should take the following form  
\be
{\cal E}^2_{\vn}\, (x,\thetat) = 
- {\pi \sigma^\prime \over 3 \lambda^2}\,  \chi\Big(\frac{\thetat \vec n}{2 \pi} \Big)+ \Big({4 \pi \sigma^\prime \over \lambda^2}\Big)^2 \, \phi^2\Big(\frac{\thetat \vec n}{2 \pi} \Big) \, x^2
\quad .
\label{efluxenconf}
\ee
The second term is the linear energy growth term whose magnitude is determined by the string tension $\sigma'$ in units $\lambda^2$. The function $\phi$ gives the k-string spectrum and it depends on the electric flux. The first term in the right hand side of Eq.~\eqref{efluxenconf} is the leading correction, related to the sometimes called L\"uscher term~\cite{Luscher:1980ac}, where the function $\chi$ is expected to be of order 1 and to  have a mild electric flux dependence.
 
It is interesting to point out that, although valid in the opposite regime than the perturbative formula, it shares some properties with it. First of all, that the formula simplifies when expressed in terms of the energy square. Indeed, the truncated formula is actually exact for the Nambu-Goto string. The second is that, once more, all the dependence comes through the two combinations $x$ and $\thetat$. Our conjecture in Ref.~\cite{Perez:2013dra} was that actually, at all values of the torus size, the energies only depend (continuously) on these two quantities. Thus, one should obtain similar energies with two very different values of $l$ and $N$ provided the product remains constant and the parameter $\thetat$ varies only slightly. This statement is stronger than the ordinary  volume independence since it is valid also at finite $N$. However, it relies upon continuity in $\thetat$. 

In the next subsection we will present the results of our non-perturbative study aiming at testing our conjecture and investigating the transition from the perturbative regime to the confinement regime. 

\subsection{Non-perturbative Results}

To compute the energies in a non-perturbative fashion we used the lattice formulation. We will first explain in very simple terms the essentials of the calculation and collect all the technical details in the appendix. 

The SU(N) model is formulated on a finite lattice of spatial size $L \times L$ and twisted boundary conditions with flux $k$. The model depends on a single coupling $b=1/(a \lambda_L) $. The lattice energies 
$E_L$ are extracted from the exponential decay in time of correlation functions of Polyakov lines with the appropriate winding and projected to the corresponding minimum  momentum value. One has still many possible operators to use, and that redundancy is used to maximize the coupling of the operator to the state whose mass one is trying to determine. The lattice quantities are all dimensionless because they are in units of the lattice spacing $a$. Dimensionless quantities are directly comparable. Hence, $E_L b = {\cal E}$ and $L/b=\lambda l$. Strictly speaking this identification has lattice artefact errors. These disappear when taking the lattice spacing to zero, implying $b$ and $L$ going to infinity with the ratio fixed. Comparing the results at different values of $b$ and $L$ one can quantify these errors. In our case, they are not expected to affect the main conclusions of the paper. We leave the detailed explanation  of the procedure and other technical aspects to the appendix and in the following we will present the results. 

\begin{table}[t]
\begin{center}
\begin{tabular}{cccccccc}
N & 5 & 7 & 11 & 13 & 17& 34 &89 \\
\hline
L & 14 & 10 & 6 &6 & 4 & 2& 1 \\
$k$ & 1-2 & 1-3& 1-5& 5& 2-8 & 13& 34\\
\hline
L & 28 &20 &  12 &12 & 8& -& -\\
$k$ & 1-2 & 1-3& 1-5& 5& 2-8 &-&-\\
\hline
L & - & -& -& -& 16 &-&-\\
$k$  & - & -& -& -&  3,5 &-&-\\
\hline
\\
\end{tabular}
\caption{Values of $N$, $L$ and $k$ used in the non-perturbative lattice simulations.}
\label{tab:params}
\end{center}
\end{table}

Our  study covers a wide range of values of $N$, $k$ and $x$. In particular, we  analyse all coprime values of $k$ for 4 different gauge groups  $N$= 5, 7, 11, 17 (cf.~Table~\ref{tab:params})  over a wide range of values of $x$  -- from the large-volume region to the small-volume perturbative one. Given that on the lattice one has $x = NL/(4\pi b)$, we proceeded as follows.  The results are divided in three groups of approximately constant value of $NL$. The coupling was then slightly adjusted to have the same set of $x$ values within each group. In the second group $L$  and $b$ are doubled and in the third one multiplied by 4. This corresponds to dividing the lattice spacing by 2 and 4, as a way to measure  the systematic lattice artefact errors. 
In addition, we also analysed  $N=13$ with $k=-\kb=5$ for a subset of values of $x$, and $N=34$ and $N=89$ (with $k=-\kb=34$ and $k=-\kb=13$ respectively) for only a few values of $x$. These last additions correspond to values of $N$ and $k$ belonging to the   Fibonacci sequence, which play a special role according to Ref.~\cite{Chamizo:2016msz} as will be explained later. 

Our results extend those obtained in Ref.~\cite{Perez:2013dra} which concentrated mostly on the minimum momentum state $|\vec n|=1$. The most important properties of our result will now be enumerated. 

\begin{figure}[t]
\centering
\includegraphics[width=.9\linewidth]{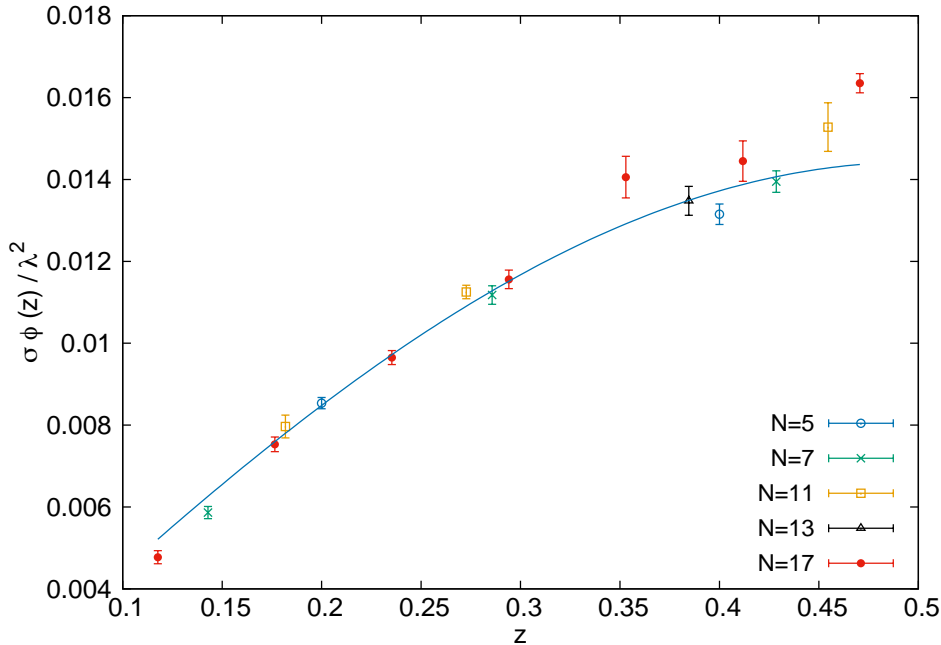} 
\captionof{figure}{We display the $z$-dependence of the function $\sigma' \phi(z)/ \lambda^2$ appearing in eq.~\eqref{efluxenconf}, extracted from fitting the $x$-dependence of the energy of electric flux $\vec e = (z N,0)$ for minimum momentum $|\vec p|= p_0$ and various values of $N$ and the magnetic flux $k$. The continuous line is the function  $ (\sigma'/\lambda^2) \sin(\pi z)/\pi$, for the best fit value $\sqrt{\sigma'}/\lambda = 0.213(1)$.}
\label{fig:sigma}
\end{figure}

\subsubsection{Consistency with expectations for small and large volumes}

At small $x$ our energies agree with the values obtained by the perturbative calculation within errors. At large $x$ they are consistent with the confinement behaviour. Indeed our calculation allows to obtain unprecedented information about the $k$-string spectrum: the dependence of the string tension on electric flux. This is encoded in the function $\phi(\thetat \vec n/(2 \pi))$ appearing in Eq.~\eqref{efluxenconf}. One can extract this function from a fit to the data to be described later.
The result is given in   Fig.~\ref{fig:sigma}, where  we display the values of the function $\sigma' \phi(\vec e/N)/ \lambda^2$  obtained from the fits compared to the function 
\be
\phi_0(\vec z) = \frac{1}{\pi}\Big|(\sin(\pi z_1), \sin(\pi z_2)) \Big|\, ,
\label{phizero}
\ee
which is one of the characteristic dependences in simple models. We also obtain a  value of the string tension  $\sqrt{\sigma'} / \lambda =0.213(1)$, which is consistent with the value 
previously obtained in Ref.~\cite{Perez:2013dra}. This can also be compared with the analytic prediction of Nair~\cite{Nair} $1/\sqrt{8 \pi}$ and the value $0.19636(12)$ obtained from a recent lattice calculation~\cite{Athenodorou:2016ebg}. 
We do not give much significance to  this $7\%$ difference,  since the large $x$ region is precisely where 
the lattice spacing corrections are expected to be larger. A detailed comparison would demand a continuum extrapolation and a thorough analysis of the systematic errors. This was not one of the main goals of this work, which covers such a wide range of sizes and values of $N$.

\subsubsection{Transition from small to large volumes}
\begin{figure}
\centering
\begin{subfigure}{.5\textwidth}
\raggedleft
  \includegraphics[width=1.\linewidth]{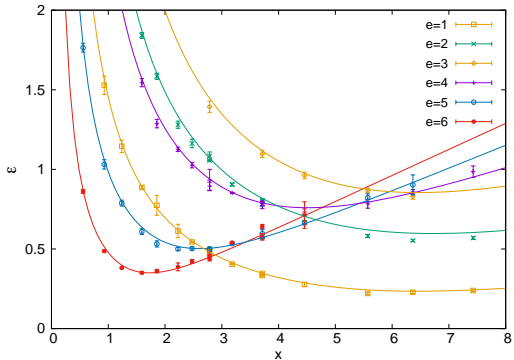}
\captionsetup{width=.9\textwidth}
  \caption{$ \{N,k\}$= $\{17,3\}$}
  \label{fig:Fig2a}
\end{subfigure}%
\begin{subfigure}{.5\textwidth}
\raggedleft
  \includegraphics[width=1.\linewidth]{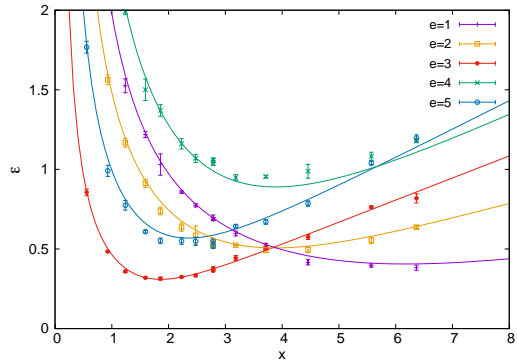}
\captionsetup{width=.9\textwidth}
  \caption{$ \{N,k\}$= $\{11,4\}$}
  \label{fig:Fig2b}
\end{subfigure}
\caption{The $x$-dependence of the torelon energies for states with electric fluxes $(e,0)$ for $ \{N,k\}$= $\{17,3\}$ (left) and $ \{N,k\}$= $\{11,4\}$ (right). The continuous lines are given by Eq.~\eqref{efluxenergy} with $\chi_0=0.6$, and $\sqrt{\sigma'} / \lambda =0.213$. }
\label{fig:Fig2}
\end{figure}

\begin{figure}
\centering
\begin{subfigure}{.5\textwidth}
\raggedleft
  \includegraphics[width=1.\linewidth]{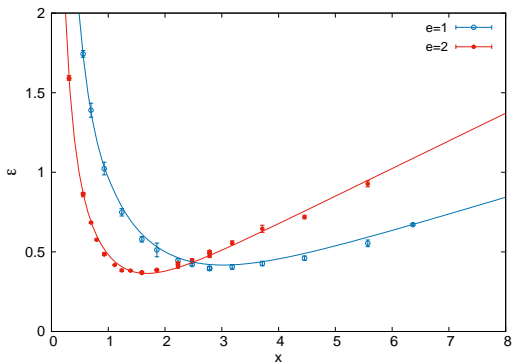}
\captionsetup{width=.9\textwidth}
  \caption{$ \{N,k\}$= $\{5,2\}$}
  \label{fig:Fiboa}
\end{subfigure}%
\begin{subfigure}{.5\textwidth}
\raggedleft
  \includegraphics[width=1.\linewidth]{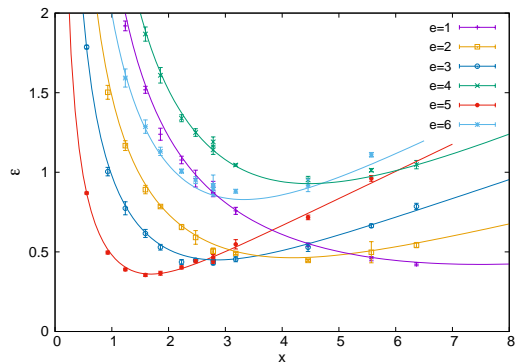}
\captionsetup{width=.9\textwidth}
  \caption{$ \{N,k\}$= $\{13,5\}$}
  \label{fig:Fibob}
\end{subfigure}
\caption{The $x$-dependence of the torelon energies for states with electric fluxes $(e,0)$, for the Fibonacci pairs $ \{N,k\}$= $\{5,2\}$ and $\{13,5\}$.  The continuous lines are determined as for Fig.~\ref{fig:Fig2}.}
\label{fig:Fibo}
\end{figure}

Our data allow us to follow the dependence of the torelon energies at all values of $x$.  The energies, that decrease as $1/x$ for small sizes, reach a minimum and then start to rise and eventually go linearly with $x$ as predicted by confinement. Except in some cases to be mentioned later, this transition is smooth. This is illustrated in Fig.~\ref{fig:Fig2} and Fig.~\ref{fig:Fibo} for different values of $N$ and $k$. 

\begin{figure}
\centering
\includegraphics[width=.9\linewidth]{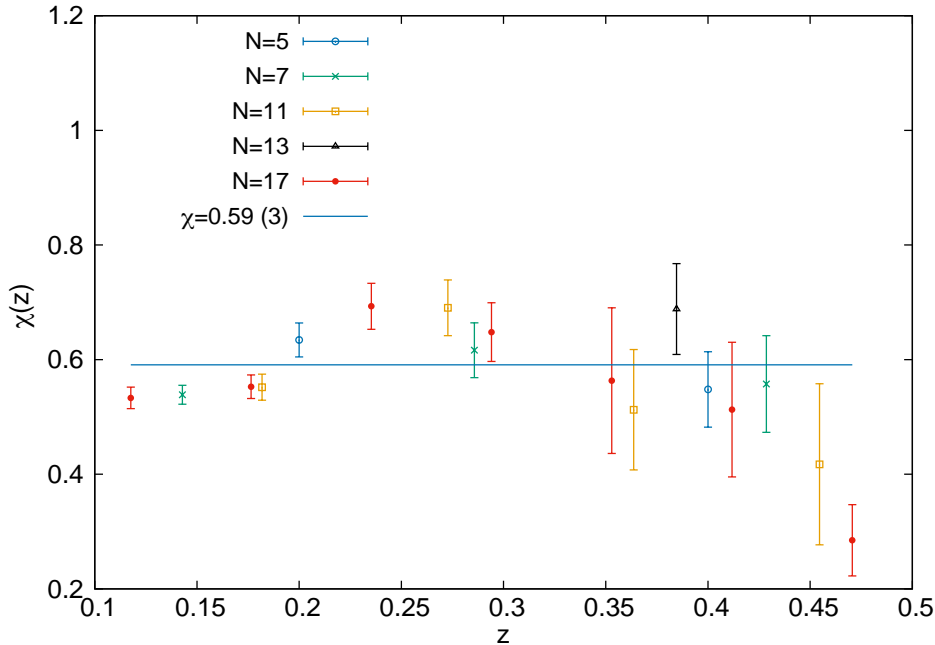}
\captionof{figure}{We display the value of the L\"uscher term function $\chi(z)$, appearing in eq.~\eqref{efluxenconf}. The $z$ dependence is extracted from fitting various $|\vec n|=1$ torelon energies to Eq.~\eqref{efluxenergy}, allowing the constant $\chi_0$ to depend on $z=\thetat/(2\pi)$ and setting $\sqrt{\sigma'} / \lambda =0.213$.}
\label{fig:chi2}
\end{figure}

The remarkable fact is that the data are well described by the continuous lines which come  from the following function 
\be
 {\cal E}^2_{\vn}\, (x,\thetat) =    { |\vn|^2 \over 4 x^2}  -   G \Big(\frac{\thetat \vec n}{2 \pi} \Big) \, {1 \over x }
- {\pi \sigma^\prime \over 3 \lambda^2}\,  \chi_0 + \Big({4 \pi \sigma^\prime \over \lambda^2}\Big)^2 \, \phi_0^2\Big(\frac{\thetat \vec n}{2 \pi} \Big) \, x^2\ ,
\label{efluxenergy}
\ee
which is obtained by simply adding the leading formulas for large and small $x$. The function $\phi_0$, given in Eq.~\eqref{phizero}, and the string tension value $\sqrt{\sigma'}/\lambda=0.213$ were explained  earlier. The L\"uscher term function $\chi(z)$ which was expected to a have  a slight electric flux dependence is fixed to a constant $\chi_0=0.6$. If one tries to do better one  can fit a different constant for each value of the argument $z$. The result is displayed in fig~\ref{fig:chi2}. In the range $z\in (0.15,0.45)$ it is consistent with a constant value $\chi(z)=0.59(3)$ with a $\chi^2$ per degree of freedom of 1.4. Hence, for our descriptive purposes it is much simpler, and hence better, to fix the value to $0.6$.

It must be mentioned that  Eq.~\eqref{efluxenergy} is not supposed to be exact. The intermediate region might have a somewhat more complex structure. Indeed, in Ref.~\cite{Perez:2013dra}  we added a term of the form
\be 
\mathcal {A} e^{-S_0/x} \frac{1}{x^3 \sqrt{x}} 
\ee
motivated by the possible contribution of sphaleron states at intermediate values of $x$. With these two extra parameters we were able to fit the data points with good $\chi^2$. This holds also for the new data and has been used to obtain the results for the $k$-string spectrum presented in Fig.~\ref{fig:sigma}. Nonetheless, since our goal is more to understand the general dependence on the arguments than to obtain a precise determination of the parameters, we prefer the simple description provided by Eq.~\eqref{efluxenergy}. In terms of only two parameters 
$\sigma'/\lambda^2$ and $\chi_0$ we are able to approximate hundreds of measured values to within a few percent.  This is clear in Figs.~\ref{fig:Fig2} and \ref{fig:Fibo} but also holds for all the other values not displayed.

\subsubsection{Continuity in $\thetat$}
Our main hypothesis formulated in  Ref.~\cite{Perez:2013dra} 
was that the energies depend on $l$, $N$ and $k$ only through the two combinations $x$ and $\thetat$. However, since the latter quantity only takes discrete values for each finite $N$, the hypothesis necessarily implies  continuity in $\thetat$. In principle, the validity of this assumption follows as a consequence of the capacity of our previous parameterization to fit the data. Nonetheless, thanks to the fact that we constructed our data at specific values of $x$ we can also  make a direct check of this continuity.

\begin{figure}[t]
\centering
\includegraphics[width=.9\linewidth]{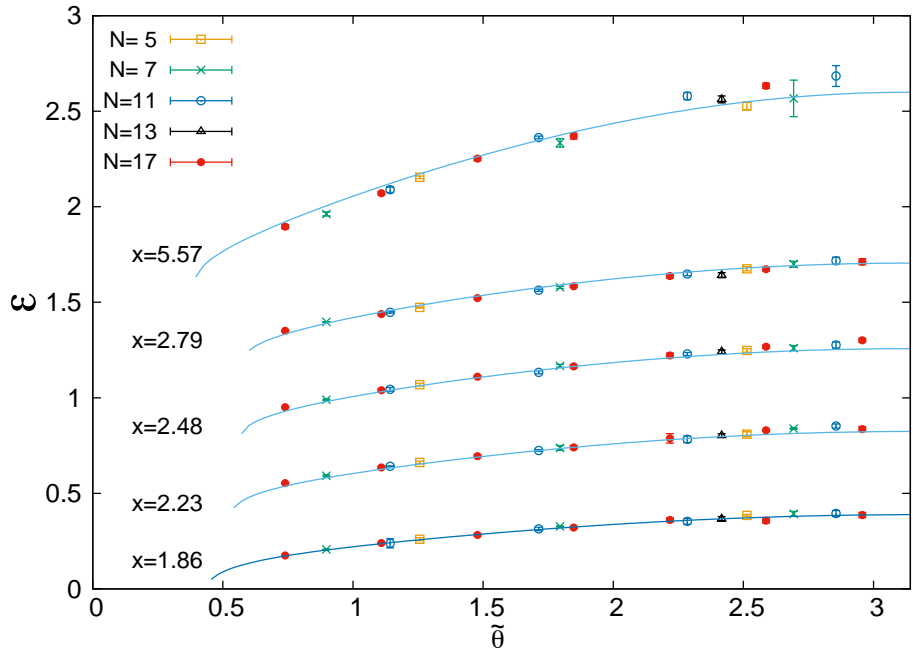}
\captionof{figure}{Dependence of ${\cal E}_{(1,0)}$ on the twist angle $\thetat$ defined in Eq.~\eqref{eq.thetat} 
for various values of $x$ and $N=5$, 7, 11, 13, 17. The results at different values of $x$
are displaced vertically by 0.4 starting from the lowest displayed value of $x$. The continuous lines represent the parameterization in Eq.~\eqref{efluxenergy} with $\chi_0=0.6$, and $\sqrt{\sigma'} / \lambda =0.213$.}
\label{fig:thetadep}
\end{figure}

In Fig.~\ref{fig:thetadep} we plot ${\cal E}_{(1,0)}$ for  
several intermediate values of $x$ and $N=5$, 7, 11, 13, 17. For the sake of clarity, the results at different values of $x$ 
are displaced vertically by 0.4 starting from the lowest displayed value of $x$. The data show a smooth dependence on $\thetat$, which is qualitatively very well described by the parameterization given by Eq.~\eqref{efluxenergy}.

\begin{figure} 
\centering
\begin{subfigure}{.5\textwidth}
  \centering
  \includegraphics[width=1.\linewidth]{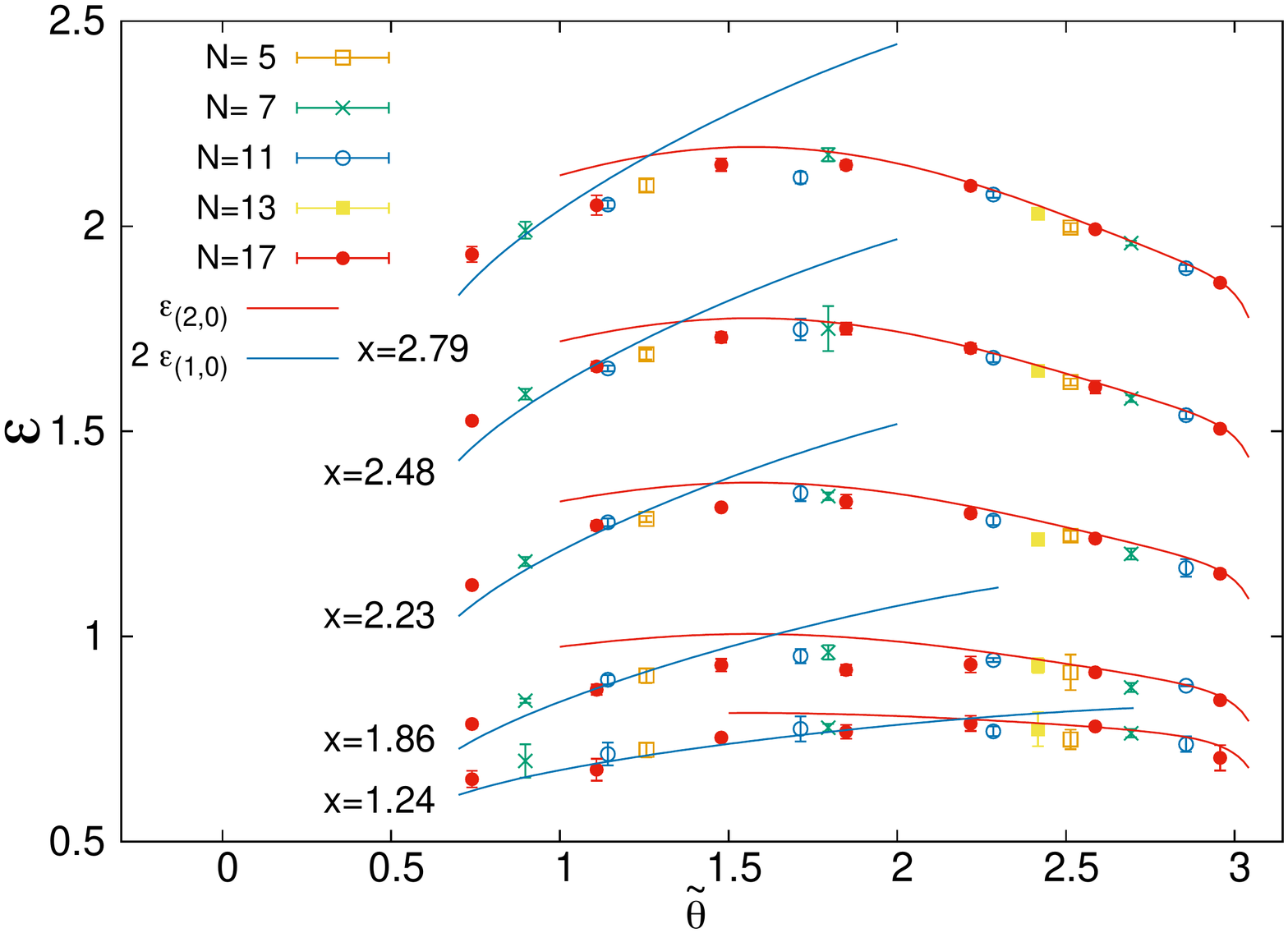}
\captionsetup{width=.9\textwidth}
  \caption{$|\vec p| =2p_0$}
  \label{fig:thetadepxa}
\end{subfigure}%
\begin{subfigure}{.5\textwidth}
  \centering
 \includegraphics[width=1.\linewidth]{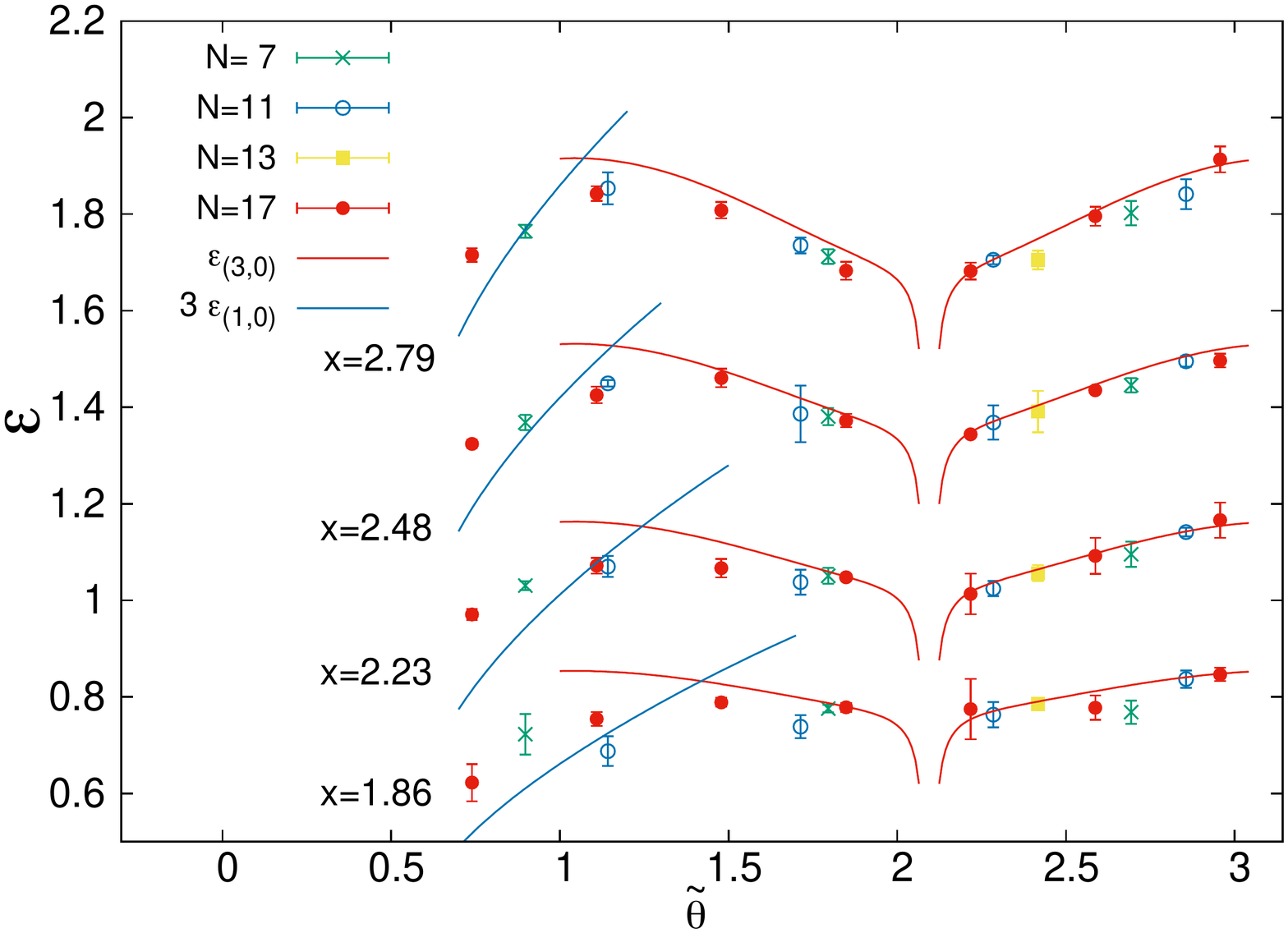}
\captionsetup{width=.9\textwidth}
  \caption{$|\vec p| =3p_0$}
  \label{fig:thetadepxb}
\end{subfigure}
\caption{Dependence of ${\cal E}_{(n,0)}$   on $\thetat$,  for states
with momentum $|\vec p| =2 \pz$ and $3 \pz$ at several values of $x$.
The results at different values of $x$
are displaced vertically by 0.4 starting from the lowest displayed value of $x$.
The continuous lines in the plot represent ${\cal E}_{(n,0)}$ or $n {\cal E}_{(1,0)}$ with ${\cal E}_{\vec n}$ given by Eq.~\eqref{efluxenergy} with $\chi_0=0.6$, and $\sqrt{\sigma'} / \lambda =0.213$.}
\label{fig:thetadepx}
\end{figure}

When testing higher values of the momenta at small torus sizes a complication arises due to the existence of degenerate  states at lowest order. The simplest situation occurs for $\vec{p}=(2\pz,0)$. The self-energy corrections are different for each state. A similar ambiguity appears at large volumes. We can have either a single string carrying the full electric flux or two strings with electric fluxes that sum up to the total one. The numerical method will just select the minimal energy of the two. Which one is smaller might depend on the value of $\thetat$.  Indeed,  that seems to be  happening according to our data. This is illustrated in  Figs.~\ref{fig:thetadepxa},~\ref{fig:thetadepxb}, where we show the dependence of ${\cal E}_{(n,0)}$ on $\thetat$,  for states with momentum $|\vec p| =2 \pz$ and $3 \pz$ respectively, at several values of $x$. For clarity, the results at different values of $x$ are displaced vertically by 0.4 starting from the lowest $x$-value. The continuous lines correspond to the predictions of our simple parameterization Eq.~\eqref{efluxenergy} for $\cE_{(n,0)}$ and $n \cE_{(1,0)}$. The data points indicate the presence of level crossings, but it is remarkable that our simple formula is able to predict where these crossings will appear. As we will see this simple exercise is a prelude of the full analysis done in the following subsection.

\subsubsection{(Non-)Existence of tachyonic instabilities }

\begin{table}[h]
\begin{center}
\begin{tabular}{ccccccc}
$N$ &  $L$ & $k$ &  $\bar{n}$ & e &  Eq.~\eqref{efluxenergy} &  $\mathcal{E}$ \\
\hline
13 & 6 & 5 & 1 & 5 & 0.53096 & 0.548(29) \\
13 & 6 & 5 & 2 & 3 & 0.45612 & 0.454(16) \\
13 & 6 & 5 & 3 & 2 & 0.50259 & 0.489(6) \\
13 & 6 & 5 & 5 & 1 & 0.76263 & 0.759(21) \\
34 & 2 & 13 & 1 & 13 & 0.52945 & 0.644(27) \\
34 & 2 & 13 & 2 & 8 & 0.46143 & 0.468(10) \\
34 & 2 & 13 & 3 & 5 & 0.49622 & 0.497(15) \\
34 & 2 & 13 & 5 & 3 & 0.76863 & 0.790(14) \\
89 & 1 & 34 & 1 & 34 & 0.52922 & 0.645(34) \\
89 & 1 & 34 & 2 & 21 & 0.46220 & 0.474(20) \\
89 & 1 & 34 & 3 & 13 & 0.49530 & 0.481(11) \\
89 & 1 & 34 & 5 & 8 & 0.75357 & 0.730(20) \\

\hline
\\
\end{tabular}
\caption{Torelon energies for various values of $N$ and $k$ belonging to the Fibonacci sequence. We show the values obtained in non-perturbative lattice simulations compared with the expectations from Eq.~\eqref{efluxenergy}.}
\label{tab:fibo}
\end{center}
\end{table}

As mentioned in the introduction the negative value of the self-energy correction might lead to the condensation of some of the electric fluxes. This happens when the energies cross zero. In our previous publications~\cite{Perez:2013dra,Perez:2014sqa} we showed cases in which this actually happens and cases in which it doesn't. Typically, the risk is higher for larger values of $N$ and smaller values of the flux $|k|$ or $|\kb|$. This fits nicely with the general proposal made in Ref.~\cite{GonzalezArroyo:2010ss} to avoid symmetry breaking in the 4 dimensional model. However, in our case it turns out that the approach of the energies to zero can be directly deduced from the approximate parameterization of Eq.~\eqref{efluxenergy}. The formula then provides a method to indicate whether there will be condensation of any electric flux torelon or not.  This works perfectly for all of our data set, which covers a wide range of values of $N$, $k$ and $\lambda$, but also allows a theoretical analysis of the presence or absence of tachyonic instabilities  for arbitrary values of $N$ and $k$. This was done in Ref.~\cite{Chamizo:2016msz}. Essentially, in order to enforce the absence of tachyonic instabilities one should guarantee that 
\be 
\min_{x, \vec n} {\cal E}_{\vec n}(x,\thetat) > 0 .
\ee
Thanks to our formula this translates into 
\be
\label{condk}
\zmin(N,k)\equiv \min_{e \perp N} e\,  \Big|\Big|\frac{k e}{N}\Big|\Big| \gtrsim 0.1 ,   
\ee
where the symbol $e \perp N$ indicate that the two integers are coprime. Is it possible for any $N$ to choose a flux $k$ that guarantees this condition is met? In Ref.~\cite{Chamizo:2016msz} it was shown that this question translates into a still  unproven conjecture in Number Theory, called the Zaremba conjecture~\cite{zaremba}. Fortunately, it has been proven recently~\cite{Huang2015} that Eq.~\eqref{condk} holds for almost all values of $N$. The analysis also shows that the largest values of  $\zmin(N,k)$, and hence of the minimum energy, occur for $N=F_n$, a member of the Fibonacci sequence, and $k=|\kb|=F_{n-2}$, for any value of $n$. Motivated by this result we included the simulation at $N=13$ and $k=-\kb=5$ which satisfies the criteria of maximal $\zmin(N,k)$. The corresponding energies are plotted in Fig.~\ref{fig:Fibob}. Next to it we have plotted the energies for $N=5$ and $k=2$, another Fibonacci optimal case. Notice the resemblance, and the ability of our simple parameterization to describe both data. 

Hence, we decided at the very late stage of this project to make some test runs at even larger values of $N$. In particular we studied $N=34$ for $L=2$ and $k=|\kb|=13$ and $N=89$, $L=1$ (only a 1 point spatial lattice) and $k=|\kb|=34$, which belong to the Fibonacci sequence.  We only studied few values of $x$ and with limited statistics. Furthermore, the code is different and we include less operators in the game. Nevertheless, the results came out to be rather good. In particular, we looked at the value $x=3.183$ which was studied for all the other values of $N$. The time correlators  of Polyakov lines show a clear exponential fall-off at moderate separations, which allowed us to obtain masses. In Table~\ref{tab:fibo}  we show the values obtained compared with the expectations from our simple formula Eq.~\ref{efluxenergy}. For completeness we also add the results for $N=13$. Notice, the good agreement specially for the lighter states. The bigger value for $\bar{n}=1$ can be due to contamination with excited states due to the limitation in number of operators.

We will study some consequences of these findings in the next subsection.

\subsection{Derived Consequences}
In this subsection we will extract several conclusions from the results presented in the previous subsection. 

\subsubsection{Conditions on the flux for the absence of phase transitions}
As we saw in the previous section it seems possible (except perhaps for a set of exceptional $N$ values) to choose the flux $k$ such that the torelon energies are always bounded from below. This guarantees the possibility of taking the large-$N$ limit avoiding any transitions as we move from small to large volumes of space. The condition is Eq.~\eqref{condk} which is stronger than the previously defined $\frac{|k|}{N}, \frac{|\kb|}{N} \gtrsim 0.1$. Thus, our proposal made in Ref.~\cite{GonzalezArroyo:2010ss} might have to be modified accordingly. Nonetheless, the first prime number $N$ in which the previous condition does not suffice is $N=61$ (with $k=21$) and the first non-prime $N=44$ and $k=21$.  Notice that the pattern suggests to remain far from a simple fraction i.e. $\frac{1}{2}$, $\frac{1}{3}$, etc. Although our new condition has been obtained for the 2+1 dimensional system it might also affect the 4 dimensional case, but replacing $N$ by $\sqrt{N}$. Indeed, it can be shown  that the  value of $\zmin$ controls the size of the contribution of non-planar diagrams in perturbation theory~\cite{Perez:2017jyq}. However, if that also affects the existence of symmetry-breaking phase transitions in the non-perturbative region of the 3+1 system would only show up for $N=44^2=1936$ or higher, which is hard to check. To avoid these problems, choosing $k$ such that $\zmin(N,k)$ is large enough is recommended. 

\subsubsection{Implications for non-commutative field theory}
As mentioned in the introduction, the system that we are studying is a particular case of a 
gauge theory on the non-commutative torus of size proportional to $x$ and non-commutativity parameter proportional to $x^2 \thetat$. The quantity $\thetat/(2\pi)$ appears as a dimensionless non-commutativity parameter which takes rational values. Since the present theory is perfectly well defined both perturbatively and non-perturbatively, this would provide a window into the non-perturbative dynamics of non-commutative field theories.  But is it possible to use this theory to investigate non-rational values of the dimensionless non-commutativity parameter?
This is one of the questions raised in an interesting paper by Alvarez-Gaum\'e and Barb\'on~\cite{AlvarezGaume:2001tv} (see also~\cite{Elitzur:2000ps})  where they also analysed in detail the possible limits that can be taken and their interpretation. They proposed to define the theory for irrational values of $\thetat$  by taking sequences of rational numbers converging to it:
\be 
\lim_{i\longrightarrow \infty}\frac{\kb_i}{N_i} = \frac{\thetat}{2\pi}.
\ee
Using the new information that we have gathered in this work and in Ref.~\cite{Chamizo:2016msz}, we posed ourselves the question if it is possible to take the limit avoiding the presence of tachyonic instability for all intermediate values $\{N_i,\kb_i\}$. 
There is at least one case in which this is possible $\thetat/(2\pi)=\frac{3-\sqrt{5}}{2}$. This is the limit of the ratio $\kb_i/N_i=F_{i-2}/F_i$, where $F_i$ is the $i$-th Fibonacci number.  However, it turns out that the set of irrational $\thetat$ for which this is possible forms an uncountable set of measure zero and non-integer Haussdorff dimension: a Cantor  set. Let us explain how this comes about.

Given the pair of coprime integers ${N,\kb}$, one can form the rational $\kb/N$. As any other rational number it admits a finite continued fraction representation 
\be
\frac{\kb}{N}=[a_0; a_1,a_2,\dots, a_M]
 :=
 a_0+1/\Big(a_1+1/\big(a_2+1/(a_3+\dots)\big)\Big),
\ee
where the so-called partial quotients $a_i$ are positive integers. Let us now call $\amax(N,\kb)=\max_i a_i$. In Ref.~\cite{Chamizo:2016msz} it is proven that 
\be
\frac{1}{\amax+2 } < \zmin < \frac{1}{\amax} .
\ee
Our condition on the absence of tachyonic instabilities (Eq.~\eqref{condk}) then imposes $\amax < 10$. The possible limiting irrationals $\thetat$ should have an infinite continued fraction built from an {\em alphabet} of only 9 digits. In this case it can be seen (see  Ref.~\cite{Huang2015}) that the set is uncountable, has zero measure and a Haussdorff dimension between 0 an 1.    

The conclusion is then quite striking. Only for a zero-measure set of values of $\thetat$ can one define a limiting theory without tachyonic instabilities. This is so  since as $N$ increases new light states appear in the spectrum that might cause a phase transition.

\section{The glueball spectrum (zero electric flux)}
\label{s.glueball}
\subsection{General considerations}
Yang-Mills theory is expected to have a mass gap in the large-volume limit and a whole spectrum of states, which are generically called glueballs. The masses of these states will be finite in $\lambda$ units.  In that limit the torelon states become infinitely  massive and decouple from the theory. The value of the glueball masses  are expected to remain finite when taking the large-$N$ limit, defining the spectrum of the large $N$ infinite volume theory.

If we now consider space to be  a very large torus with twisted boundary conditions, these glueball states  will still be there and with masses that are almost  insensitive to the torus size $l$ and the magnetic flux $k$. However, there will also be  other states in the zero electric flux sector which are made of torelon pairs with masses proportional to the size of the torus $l$. 

As we reduce the torus size the hierarchy between torelon pairs and glueballs  gets reduced and there could even be mixing between them. Indeed, in the limit of very small volume the torelons are just single gluons and the lowest energy states in the zero electric flux sector are these torelon pairs. Making the volume slightly larger these masses get corrections which, within that domain, will continue to depend jointly on $lN$ (actually on $x$) and also on $\thetat$.
Our calculation at next to leading order of perturbation theory gives modifications to the masses of these two-gluon states coming from the self-energy of the individual gluons, plus an interaction term that splits the masses of  the states into those that transform differently  under the  discrete rotation group. The state with the lowest energy is the rotationally invariant one and the interaction energy equals~\cite{Perez:2013dra}
\be
{\cal E}_{\rm int} = - \frac{3 }{4 \pi^2} \sin^2 (\thetat/2)\ , 
\ee
written in terms of the twist angle $\thetat$.

How does the transition from small to large volumes take place? Is there any type of volume independence present in this spectrum? In principle the answer to this last question seems to be no. As we saw in the previous section as the volume becomes larger the lowest-lying torelon spectrum is far from being volume independent.
When $N$ grows new light states appear and these have energies that go with $l$, not $lN$. The corresponding zero-electric flux  pairs will also have energies that grow with $l$. Thus, the statement that the spectrum of the theory at large $N$ does not depend on the volume does not hold. At $x$ fixed with low  $l$ and large $N$, some  torelon pair states will remain in the low-lying spectrum as opposed to the infinite volume case. These considerations serve to prepare for the presentation of our non-perturbative results interpolating between both regimes.

\subsection{Non-perturbative results}
\label{s.results_g}

In this section we will present our results about the spectrum of the theory in the sector of vanishing electric flux. The methodology is similar to the one used for torelon states. Masses are extracted from the exponential decay in time of correlators between operators carrying no electric flux and projected to zero momentum. 
The spectrum is analysed using the GEVP method (see appendix)
applied to the basis of operators described in sec.~\ref{s.determination}. The operators are of two types: either ordinary single trace Wilson loops with no winding on the torus, or products of two single trace Polyakov loops carrying opposite values of the electric flux. Then main features of the result will be explained now.  The actual values for the mass gap and the next excited state energy are collected in tables~\ref{tab:n5_glue} -~\ref{tab:n17c_glue} in Appendix~\ref{s.tablesg}. 

\subsubsection{$x$-dependence of the mass gap in the $\vec e= 0$ sector}

\begin{figure}
\centering
\includegraphics[width=.9\linewidth]{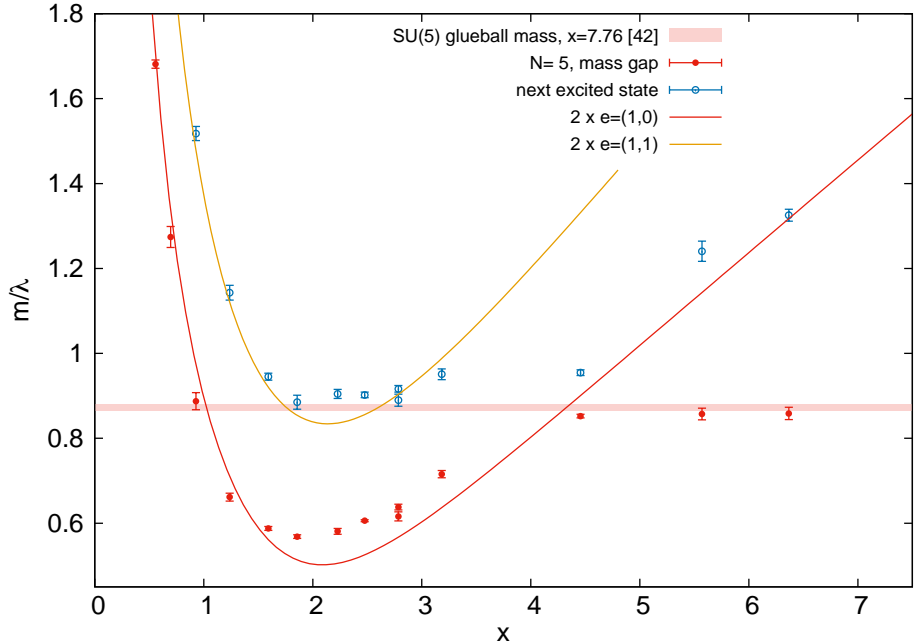}
\captionof{figure}{$x$-dependence of the mass gap and next excited state energy in the zero-electric flux sector for:  $\{N, k\} = \{5,1\}$.
The red band is the $SU(5)$ result
of Ref.~\cite{Teper:1998te}:
 $m/\lambda= 0.873(8)$, obtained on a $32^3$ lattice at $x=7.76$.
The continuous lines correspond to twice the torelon energies for the indicated electric fluxes according to our formula \eqref{efluxenergy}.}
\label{fig:gluegst}
\end{figure}

\begin{figure}
\centering
\includegraphics[width=.9\linewidth]{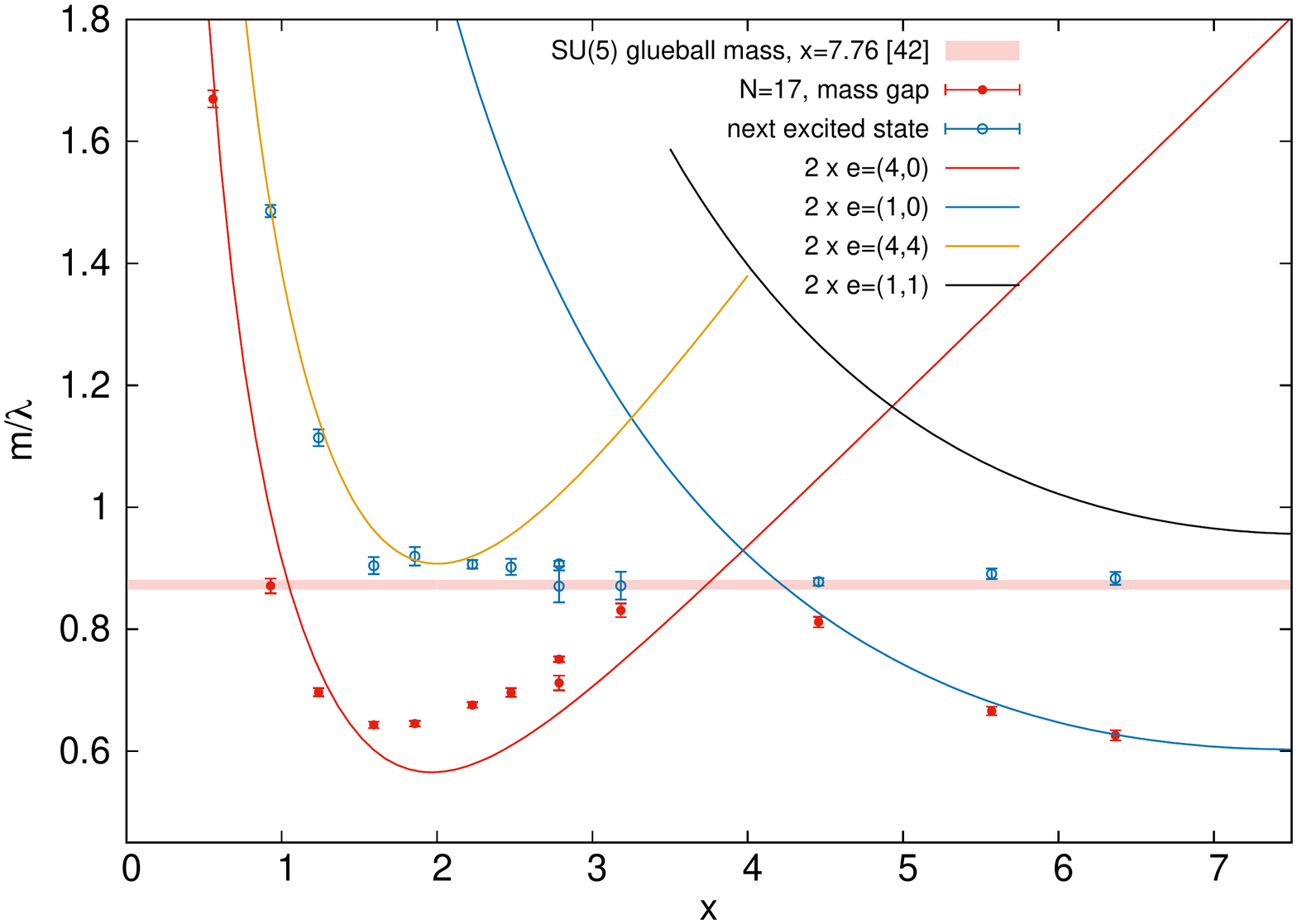}
\captionof{figure}{$x$-dependence of the mass gap and next excited state energy in the zero-electric flux sector for:
$\{N, k\}= \{17,4\}$. The red band is the $SU(5)$ result
of Ref.~\cite{Teper:1998te}, on a $32^3$ lattice at $x=7.76$. The other lines are twice the predicted torelon mass for the indicated electric flux.}
\label{fig:gluegst17} 
\end{figure}

Let us illustrate the $x$-dependence by focusing in two particularly neat examples, those corresponding to $\{N, k\} = \{5,1\}$ 
and $ \{17,4\}$, which have relatively  close values of $\thetat$ equal to 1.257  and  1.478 respectively. 
The results for the mass gap and next excited state are  displayed in Fig.~\ref{fig:gluegst} and \ref{fig:gluegst17}. Let us focus first on the dependence for the mass gap. For $x$ smaller than 1 or 2 it  follows the characteristic $1/x$ behaviour predicted in perturbation theory. Indeed, 
the  mass is quite close to twice the minimum torelon mass ($|\vec e|=|\kbar|$), indicating that this state is actually a torelon pair state. To see this, we display in the figure the lines obtained by multiplying by 2 the simple parameterization  Eq.~\eqref{efluxenergy} of the corresponding torelon. 
The difference between the double torelon mass and the mass gap is a measure of the torelon-antitorelon interaction energy. At higher values of $x$ the mass gap has a minimum in the interval $[1.5,2]$ and then starts to rise. If we first focus on  the $N=5$ case  (Fig.~\ref{fig:gluegst}) we see that  beyond $x\sim 4$ the mass seems to level up and tends to a constant. The corresponding state  in this larger $x$ region would correspond to the true lowest mass glueball which is there in the infinite volume limit. Indeed, the value of the mass is rather compatible  with the value $m/\lambda= 0.873(8)$ given in Ref.~\cite{Teper:1998te}  obtained for $N=5$ on a $32^3$ lattice at $x=7.76$. This value is indicated by the horizontal red band in the figure. Hence, the $N=5$ result is as expected and indicates a change of nature of the lowest mass state somewhere in the interval $x\in [3.5,4.5]$. This change occurs when the mass of the torelon pair becomes higher than the mass of an initially  more massive state which ultimately becomes the glueball. This is clear when looking at the energy of the next excited state, which beyond $x\sim 4$ seems to extend  nicely the behaviour of the lower $x$ mass gap with the  characteristic linear growth of a torelon-antitorelon pair. In the figure we also show the curve corresponding to twice the energy of the torelon with electric flux $\vec{e}=(|\kbar|,|\kbar|)$, which, consistently with expectations, describes the behaviour of the excited state at low values of $x$.

\begin{figure}
\centering
\includegraphics[width=.9\linewidth]{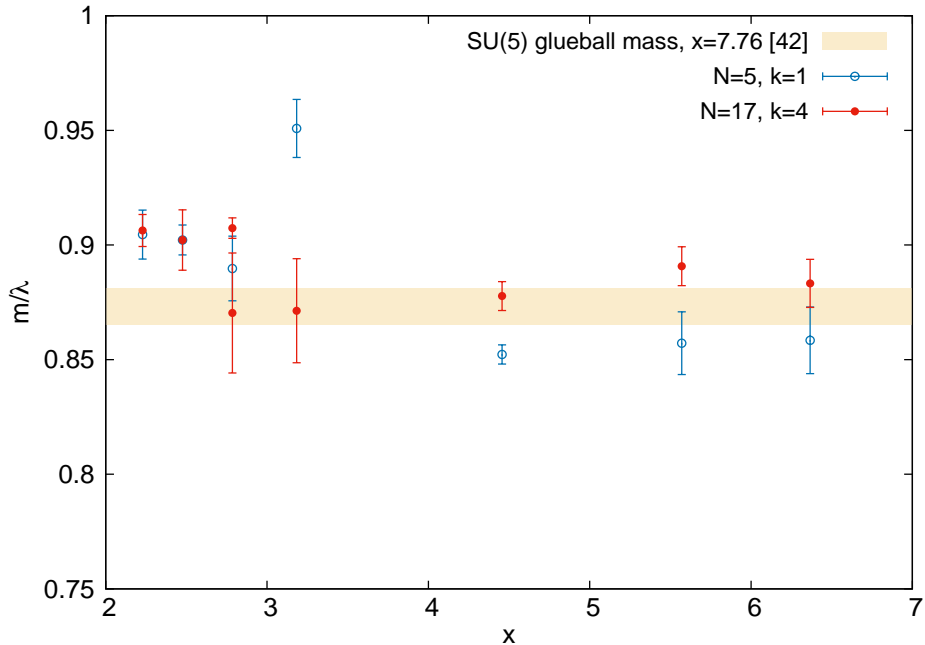}
\captionsetup{width=.9\textwidth}
\captionof{figure}{Comparison between the $x$ dependence of the masses of the glueball states
in $\{N, k\} = \{5,1\}$ and $ \{17,4\}$.}
\label{fig:glue5to17}
\end{figure}

If we now focus on the $N=17$ data (Fig.~\ref{fig:gluegst17}) we see that the behaviour of the mass gap and the next excited state for  small $x$ follow the same pattern as for $N=5$, making clear its torelon-antitorelon nature. However, the behaviour of the mass gap for  $x>4$ is quite different to the $N=5$ case, since the mass is actually decreasing in that region. The interpretation becomes obvious once we compare the data points with the blue line which is the predicted behaviour for a pair of torelons with opposite electric fluxes $|\vec e|=1$.  This was the expected trouble for larger values of $N$ arising due to the appearance of new light torelon pair states.

One may wonder if this result implies a failure of $x$-scaling in the glueball spectrum. We will argue this is not the case. For that purpose we have to look at the
next excited state for the  $SU(17)$ glueball sector. We see on Fig.~\ref{fig:gluegst17} that beyond $x\sim 3$  its mass 
approaches the value corresponding to the infinite volume glueball, with very little $x$-dependence.  These results indicate that, although the $SU(17)$ {\it  glueball} is not the lowest excited state in the zero-electric-flux sector, it is present in the spectrum for $x\gtrsim 4$ with a mass compatible to that of the $SU(5)$ glueball -- a blow-up of the comparison in Fig.~\ref{fig:gluegst17} for the large $x$ regime is presented in Fig.~\ref{fig:glue5to17}. For completeness, we also show in Fig.~\ref{fig:gluegst17} that the predicted next excited torelon pair, depicted by the black curve,  would have a higher mass than the bona fide glueball.   

Summarizing, the onset of the would-be {\it large-volume glueball} takes place in the same range of $x$-values for both $N=17$ and $N=5$. Note that $x=4$ corresponds to $\lambda l = 0.8 $ for $SU(5)$ and a much smaller physical volume $\lambda l = 0.24$ for $SU(17)$. We conclude hence that the quantity setting the onset scale for the appearance of the large-volume glueball is $x$ and not the physical volume of the box. 

Notice that this opens the door for the possibility of extracting a glueball spectra in the large-$N$ limit at fixed values of $x$. This is in line with our $x$-scaling hypothesis translated to the glueball spectrum, which is a stronger statement than volume independence. Taking $N$ to infinity at fixed $x$ implies that the size of the torus goes to zero. This is one of the limits conceived in Ref.~\cite{AlvarezGaume:2001tv} and called singular by their authors.

In sec.~\ref{s.glueselect} we will discuss how to optimize the selection of the large-volume glueball by looking to the overlap of the zero-flux states onto two-torelon and
Wilson loop operators. We will present evidence that the onset of the large-volume glueball has taken place for $x\gtrsim 4$ in all the cases we have analysed.  

\begin{figure}[tp]
\centering
\includegraphics[width=.9\linewidth]{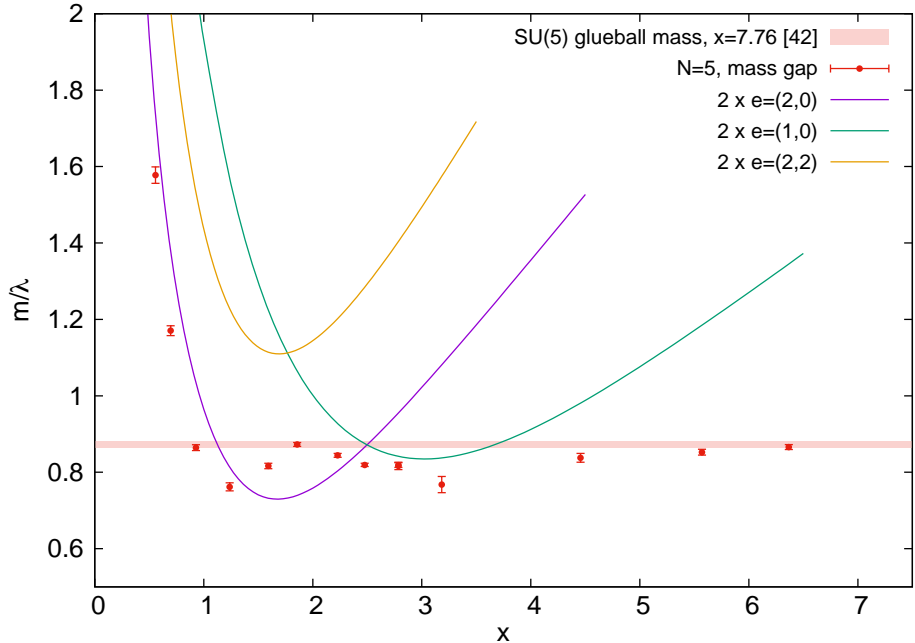}
\captionof{figure}{ $x$-dependence of the mass gap in the zero-electric flux sector for the Fibonacci set $\{N, k\} = \{5,2\}$. The continuous lines give twice the torelon energy in the corresponding electric flux sector. 
}
\label{fig:gluegst_fibo}
\end{figure}

\begin{figure}[tp]
\centering
\includegraphics[width=.9\linewidth]{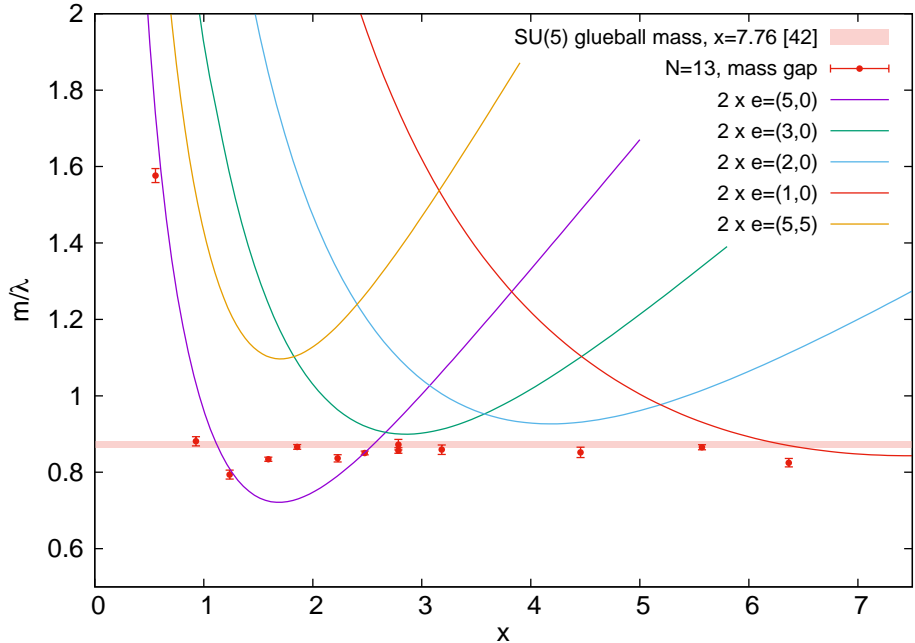}
\captionof{figure}{$x$-dependence of the mass gap in the zero-electric flux sector for the Fibonacci set $\{N, k\} =  \{13,5\}$. The continuous lines give twice the torelon energy in the corresponding electric flux sector.
 }
\label{fig:gluetor_fibo}
\end{figure}

To close this section, let us  mention that the Fibonacci sets are also optimal from the point of view of determining the true glueball mass. This is illustrated in Figs.~\ref{fig:gluegst_fibo} and \ref{fig:gluetor_fibo} where, as before, we look at the $x$-dependence of the mass gap compared with the two-torelon energies for $\{N, k\} = \{5,2\}$ and $ \{13,5\}$. For the particular cases presented here it turns out that the glueball is the  lowest energy state for a large region of $x$ values even for the larger $N$. The decrease at large $x$ for $N=13$ is in line with the presence of the $|\vec e|=1$ torelon-antitorelon pair. 
Notice that the minimum energies of the pairs are larger for the Fibonacci set as expected.  We will now explain a very interesting behaviour for the torelon energies of this  class of models.  

Given $N=F_n$ and $k=|\kbar|=F_{n-2}$, one can see that the electric fluxes that give minimum energies at all intermediate values of $x$ are precisely those corresponding to other Fibonacci numbers (this fact was used in writing Table~\ref{tab:fibo}). Furthermore if $e=F_s$ the corresponding value of the minimum momentum is given by $\bar{n}=F_{n-s}$. 
This follows from the identity (exercise for the reader)
\be
F_s F_{n-2} =F_n F_{s-2} +(-1)^s F_{n-s}.
\ee
Then by definition
\be
\zmin= \min_e \frac{e  \bar{n}}{N}= \min_s \frac{F_s F_{n-s}}{F_n}=\frac{F_{n-2}}{F_n}=\frac{|\kbar|}{N}.
\ee
For large $n$ the value quickly approaches $\frac{3-\sqrt{5}}{2}\sim 0.381966$. However, the corresponding numbers for other values of $s$ lie always between this value and $1/\sqrt{5}\sim 0.44721$. This fact implies that the minimum energies for the fluxes lying at all intermediate values of $x$ are all very close to each other. This can be seen in Fig.~\ref{fig:gluetor_fibo} where when moving from right to left the minimum fluxes are 1, 2, 3, 5. In the Figure we actually plot twice the torelon energy, and the minimum is always quite close to the value of the infinite volume glueball. It is tempting to think that this behaviour is not by chance.

\begin{figure}[t]
\centering
\includegraphics[width=.9\linewidth]{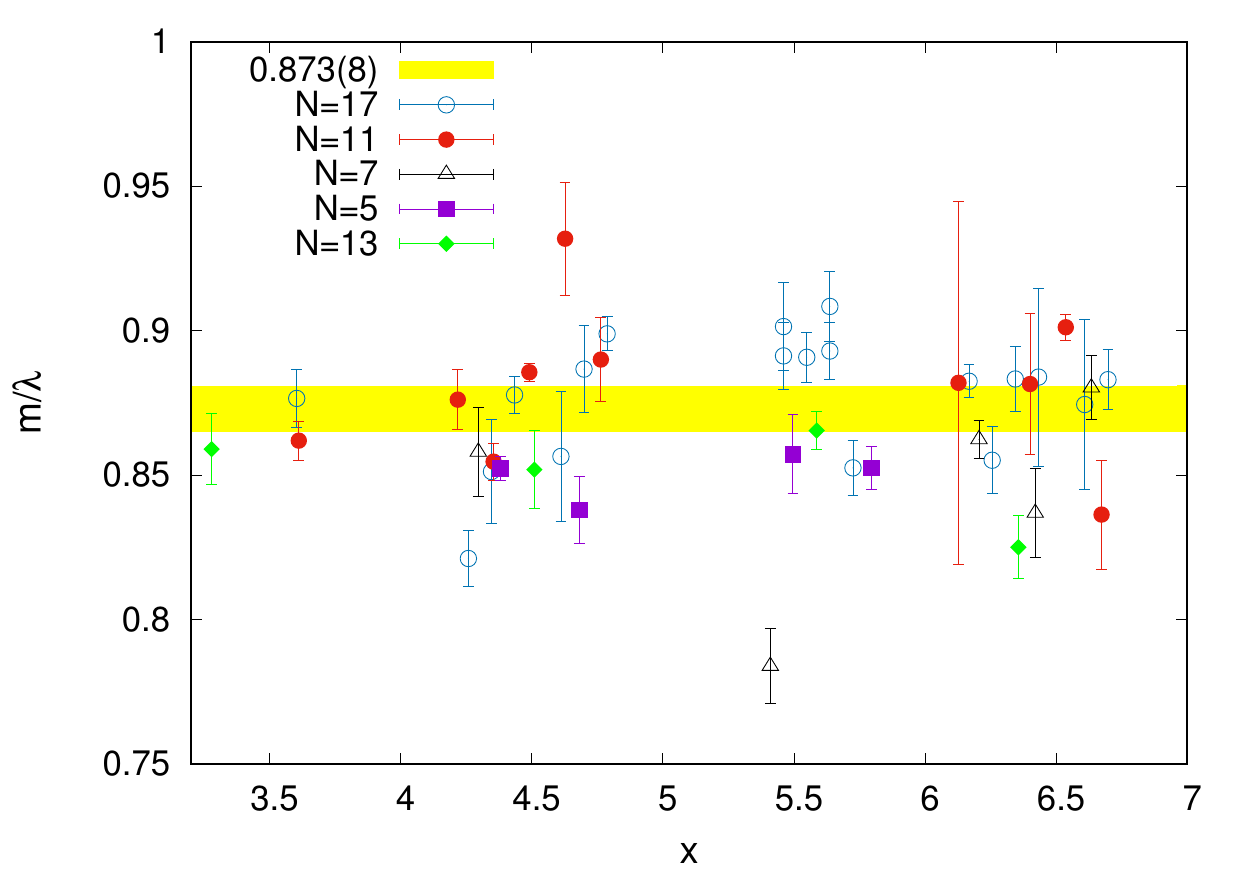}
\captionof{figure}{Large $x$-dependence of the glueball mass computed with the selection criterion discussed
in sec.~\ref{s.glueselect}.}
\label{fig:glueallN}
\end{figure}

\subsubsection{Filtering out the glueball from the different torelon pairs}
\label{s.glueselect}

Our previous subsubsection shows, for $x>4$ at least, the presence of a state in the spectrum with approximately the same mass as the infinite  volume glueball. This happens for most  values of $N$ and $k$ in our data set. The problem is that this state might not be the minimum energy  state (above the vacuum).  The question that we are trying to answer in this section is whether this state has some characteristics that allows to single it out from the remaining states in the low-lying spectra. A priori one would expect that the torelon pair states would couple more to the operators involving a product of the corresponding Polyakov lines. On the contrary, this glueball state would show preference for the Wilson loop operators. With this idea in mind we have analysed the spectra in the vanishing electric flux sector. Our data only allow to determine the mass of the two lowest mass states. For each state of our $N=5$,$7$,$11$,$17$ data we computed the projection onto a certain restricted  set of operators (the technical aspects of this procedure are explained in the appendix) and selected those states for which  the projection onto Wilson loop operators is larger than 0.7. The resulting masses are displayed in  Fig.~\ref{fig:glueallN} as a function of $x$. Since many of  the data points have the same value of $x$ we have slightly displaced the data horizontally by a quantity proportional to $\thetat$. Notice the large number of  points coming from all values of our parameters and the relatively small spread centred around a mean value of $0.878$. This value matches perfectly with   the  $SU(5)$ lattice glueball mass from Ref.~\cite{Teper:1998te} presented in previous plots, which appears as the yellow band in the figure. Our procedure does not work for our $N=13$ data points, presumably because of the presence of almost degenerate torelon pair states. The $N=13$ data points in the figure are just the value of the mass gap. The mass degeneracy is quite obvious. In units of the string tension, we obtain $m/\sqrt{\sigma}= 4.10(4)$, compatible within errors with the  large 
$N$ glueball mass as estimated in Ref.~\cite{Athenodorou:2016ebg}. This provides a confirmation that beyond $x\sim 3.5$ the glueball state is present in the spectrum.

\subsection{Derived Consequences}
As we did in the previous section with the torelon spectra, we will explain in this subsection the consequences that one can extract from the results presented in the previous one. 

\subsubsection{$x$-scaling and  volume independence in the $\vec e=0$ sector}
Our results show a spectra in agreement with  expectations. In the range of $x$ values studied  there are states  with masses approximately given by the sum of two torelon ones, as given by our simple formula. They correspond to torelon pairs. It makes sense that for large $N$  the mass of a torelon pair comes close to the sum of the masses. Interaction energies can be obtained from our data by taking the difference.  

Our data shows also a state coupled mostly to Wilson loops and with a mass which is close to the one of the infinite volume glueball. This state can be identified with the glueball. It  appears in the low-lying spectrum at around a similar value of $x\sim 3$ independently of the other parameters, despite the large difference  in values of $l$ and $N$ involved. This implies that   some sort of $x$-scaling is actually in play. However, the data  does not allow to establish any pattern in the mass differences as a function of $N$ or $\thetat$ (see Fig.~\ref{fig:glueallN}). It might be due to the fact that the differences are small and can be overtaken by all sorts of systematic and statistical uncertainties. 

One can criticize  the identification of the states as torelon pairs and glueballs. Since they have similar quantum numbers these states might mix and the mixing can affect the value of the masses too. Nonetheless, the mixing is expected to be bigger for states that are nearly degenerate. Most importantly the mixing is suppressed in the large-$N$ limit and this makes the possibility of having a well defined glueball spectrum separated from a torelon pair spectrum viable even when the masses are not very different. Nevertheless, the criticism certainly applies at finite $N$ where the states will be mixed with the torelon pair states which show a strong volume dependence. 
Still this might induce small corrections for $N$ sufficiently large. What could happen in the large-$N$ limit will be discussed in the next paragraph.  

Let us first clarify the possible options that occur when taking the large-$N$ limit. The more standard way would be to take the limit at fixed value of $\lambda l$. This implies $x$ would go to infinity. According to volume independence the glueball spectrum should be independent of $l$ and of $k$. Although at finite $l$ the torelon pair spectrum is still relatively light  it is decoupled from the standard glueball spectrum. Our results show that the limit can be taken avoiding the presence of tachyonic instabilities paying the price of choosing the flux $k$ appropriately for each $N$. This procedure is in line with the way in which one obtains results in 4 dimensions using the twisted Eguchi-Kawai model. Even with wrong choices of the flux the limit can still be taken for values of $\lambda l$ which are sufficiently large to lie beyond the region of instability. This would match nicely with the proposal of Narayanan and Neuberger~\cite{Narayanan:2003fc} using zero-flux (periodic boundary conditions). However, using the right fluxes this minimum length can be decreased at will.

A much stronger type of large-$N$ limit occurs  when it is taken at fixed value of $x$, since in that case the size is driven to zero as $N$ grows. This is the singular limit discussed in Ref.~\cite{AlvarezGaume:2001tv} and also the one involved in $x$-scaling. Our results show that talking about a glueball spectrum in that limit might  make sense at least for $x$ beyond a certain threshold value. Furthermore, the mass gap has a very mild $x$ dependence in that region. Whether this resulting spectra would depend on $\thetat$ is somewhat more questionable. We saw that the torelon spectra  only allows to take the singular limit for a zero-measure set of $\thetat$ values, without falling into a phase transition region at intermediate values of $x$. Still it could be possible that this does not affect the glueball states in question, but this is hard to defend this possibility without any type of solid argument.

\subsubsection{Beating factorization for the glueball spectrum}
Our paper has also implications in one of the main challenges of non-perturbative large-$N$ gauge theories. This has to do with the difficulty of extracting the glueball spectrum due to factorization. If one computes the correlation of two Wilson loops at different times, the leading term will be the uncorrelated term, while the correlation carrying the signal of the time dependence is suppressed as $1/N^2$. Thus, it could be completely covered by the fluctuations of the uncorrelated part. One way to beat this problem is to consider large Wilson loops. In the confinement regime the expectation value of these large Wilson loop will become rather small, while this size does not in principle affect the coupling of the loop to the glueball. If we make a very naive estimate based on the area law, one might conclude that it would be enough to take the area to grow as $\log N$ to make the correlated and uncorrelated pieces of the same order. Even for reduced models in which the loop sizes  are forced  to be much smaller than $N$ or $\sqrt{N}$  (which acts as an effective box size) this is acceptable. Our experience in the present case   shows that this is indeed possible.

\section{Conclusions}

In this paper we have analysed the dependence of the spectrum of 2+1 dimensional Yang-Mills field theory on the  volume of space. Ultimately, the goal is to have a complete understanding of the dynamics of this system, which is certainly a prerequisite before achieving the same goal for the more complex 3+1 dimensional theory. It would be nice to be able to test and substantiate analytic approaches as that followed in Ref.~\cite{Nair}. Here, we use the  spatial volume, and the boundary conditions on it as a probe, which allows us to better understand the different regimes present and the transition among them. Furthermore, we study a wide variety of values of $N$ since we expect a simplification of that dynamics for large $N$. Our analysis has many implications for several sideline problems like that of volume independence and non-commutative field theories. The main implications of our results have been laid down  at the end of the previous sections dealing with non-vanishing and vanishing values of the electric flux. We briefly summarize the main points below. 

We have been able to synthesize the evolution of the torelon (non-zero electric flux) energies from the perturbative regime to the confinement regime into a simple formula. This has allowed us to predict the conditions for the avoidance of tachyonic instabilities in the system. This can be maintained at all stages when taking the large $N$ limit, but only for a measure zero set of values of the  dimensionless non-commutativity parameter  $\thetat$. In the zero-electric flux sector we have been able to disentangle in the spectrum torelon pairs from genuine glueballs. The emergence of the latter with a relatively constant mass value  takes place at a given value  of the effective size $l N$ and hence when volumes are still small enough for the torelon pairs not to become very massive. Thus, despite the complex volume dependence of torelon spectra a largely volume independent glueball state arises.

The data contained in this work  represents a considerable effort given the large number of simulations involved. Our results cover a huge range of values of $N$ and also of sizes implying not only an important computational method but also the necessity of dealing with different technical problems in different ranges of parameters. Nonetheless, this work is certainly improvable along various directions. In our opinion, the largest room for improvement is at the larger values of $x$. This would probably demand a more complete list of operators which could also allow a better determination of the spectra, a larger number of excited states and an exploration of different quantum numbers. 

\section*{Acknowledgments}

We have profitted with many conversations with colleagues during the development of 
this work. We signal very specially Carmelo P\'erez Mart\'{i}n,  Fernando Chamizo 
and Jos\'e Fernandez Barb\'on. A. G-A wants to thank the Department of 
Theoretical Physics at Tata Institute of Fundamental Research for funding 
his stay which allowed interesting discussions with the members of the 
Department. A. G-A also wants to thank the Salvador de Madariaga program 
(Ref. PRX17/00504) of the Spanish Ministry of Education  for funding his stay at Rutgers University where the last stages of this work were completed. Special thanks go also to Herbert Neuberger for many discussions. 
M.G.P. and A.G-A acknowledge financial support from the MINECO/FEDER grant
FPA2015-68541-P  and the MINECO Centro de Excelencia Severo Ochoa Programs SEV-2012-0249 and SEV-2016-0597.
M. O. is supported by the Japanese MEXT grant No 17K05417 and the MEXT program for promoting the enhancement of research universities. 
We acknowledge the use of the Hydra cluster at IFT.

\appendix

\section{The  Lattice simulation}

In this appendix we explain the methodology used in our lattice simulation.

\subsection{Lattice model}

We analyse a lattice $SU(N)$ model in 2+1 dimensions where the spatial components are
defined on a torus with twisted boundary conditions and the temporal extent is
taken periodic but always large enough to be able to neglect the effects of finite
temperature, cf.~Sec.~\ref{s.finiteT}. The action of the model is:
\begin{equation}
S = N b \sum_n \sum_{\mu\neq\nu}\left(N-z^*_{\mu\nu}(n)\Tr \left(U_\mu(n) U_\nu(n + \hat \mu)  U_\mu^\dagger(n + \hat \nu) U_\nu^\dagger(n)\right)\right).
\end{equation}
The index $n$ goes over the sites of the $L\times L\times T$ lattice, with the physical size of the 2-torus given by $l=La$, with $a$  the lattice spacing.
The twist tensor $z_{\mu\nu}$ is equal to 1, except at one corner plaquette of
each spatial plane where it is:
\begin{equation}
z_{ij}(n) = \exp\big(i\epsilon_{ij}\frac{2\pi k}{N}\big).
\end{equation}
The dimensionless coupling $b$ is defined as $b\equiv 1/(a \lambda_L)$, with $ \lambda_L$  the 't Hooft coupling on the lattice, differing from the continuum $\lambda$ by lattice artefacts. We will express most lattice computed quantities in units of $\lambda_L$. Other choices are possible, for instance  
Ref.~\cite{Athenodorou:2016ebg} uses a mean-field improved coupling~\cite{Parisi:1980pe}. For the coarsest lattices that we are using, the difference amounts to about $10\%$ and should go to zero in the continuum limit.

\subsection{Determination of the spectrum}
\label{s.determination}

The spectrum of the non-zero electric flux states has been extracted directly from the exponential decay of the two-point correlation function of Polyakov loop operators with appropriate winding number and fixed non-zero minimal momentum allowed by the twisted boundary conditions. 
The Polyakov loops can be constructed in terms of single winding operators represented by the product of APE-smeared \cite{Albanese:1987ds} link variables
\be
P_x (t,y)  =  \prod_{s=0}^{L-1} U_1^{(s)} (t,x+s,y)\quad ,
\ee
and analogously along the $y$ direction. For the results presented in this paper we have taken a fixed number of APE smearing  steps: $s=21$. 
A generic Polyakov loop of winding $\vec w$ is given by: ${\cal P}_{\vec w} = {\rm Tr} (P_x^{w_1}P_y^{w_2})$~\footnote{For details on how to project over minimal momentum see  Ref.~\cite{Perez:2013dra}.}. In most cases we have only considered Polyakov loops winding around one torus cycle but in a few cases
 we have also analysed correlation functions of loops of the form ${\rm Tr} (P_x^{e}P_y^{e})$.

 In order to improve the overlap onto the lowest mass state for the zero electric flux sector, we use the Generalized Eigenvalue Problem (GEVP) ~\cite{Michael:1985ne,Luscher:1990ck}, that has become a standard tool to compute the spectrum in lattice gauge theories. We will briefly discuss below the particular implementation of the GEVP we have used to determine the glueball masses.

The basis of observables used to extract the spectrum in the zero electric flux sector consists of rectangular Wilson loops $W(n,m)$ and moduli of multiwinding spatial Polyakov loops $|\Tr P^n|^2$.\footnote{As in the case of non-zero electric fluxes, in some cases we also include operators of the form $|\Tr P_x^nP_y^n|^2$.} The latter capture the ``torelon-antitorelon'' states wrapping around the finite torus, while the former are most useful in the large-$x$ regime where they couple to the large-volume glueballs.

We use three fixed levels of APE smearing \cite{Albanese:1987ds} with $7,14,21$ iterations (as well as the unsmeared operators, which are however too contaminated by excited states to be used in practice). Small Wilson loops can be dominated by UV effects, therefore to maximize the overlap to the physical states, we include in the basis the rectangular Wilson loops of large extent, ranging up to approximately 20-40 lattice units, depending on the value of the coupling. This can be seen as an alternative to the blocking procedure \cite{Teper:1987wt,Teper:1987ws}.

Note however, that this set-up is somewhat limited -- it is best suited for the extraction of the lowest mass glueball, while the lack of decorated loops is expected to give at best a qualitative description of the excited glueball states.

Given the observables $O_i(t)$, the matrix of two-point connected correlation functions is given by:
 \begin{equation}
 C_{ij}(t) = \sum_{t'}\langle O_i(t'+t)O_j(t')\rangle-\langle O_i(t'+t) \rangle
 \langle O_j(t')\rangle.
 \end{equation}
A typical size of the correlation matrix we used for the glueball mass extraction is between 15 and 30 operators.

To obtain a good overlap with the lowest energy states we perform the GEVP on the correlation-function matrix \cite{Michael:1985ne,Luscher:1990ck}:
\begin{equation}
C(t_1)v^{(\alpha)}(t_0,t_1) = \lambda^{(\alpha)}(t_0,t_1)C(t_0)v^{(\alpha)}(t_0,t_1)
\label{eq:eigenvec}
\end{equation}
where the times are fixed to $t_0=a$, $t_1 = t_0+a$ and $\alpha$ labels the energy states. The obtained eigenvectors $v^{(\alpha)}(t_0,t_1)$ are then used to find the improved basis of correlation functions \cite{Michael:1985ne}:
\begin{equation}
\tilde{C}^{(\alpha \beta)}(t) = \big(v^{(\alpha)}(t_0,t_1), C(t)v^{(\beta)}(t_0,t_1)\big)
\end{equation}
for each value of $t$ and then the diagonal values  $\tilde{C}^{(\alpha \alpha)}(t)$ are used to find plateaux ranges and subsequently fit to the selected ranges\footnote{In practice we limit ourselves to the two lowest-lying states $\alpha=0,1$.}. 

Note that this is different from the approach of Ref.~\cite{Blossier:2009kd} which scales $t_0$ proportionally to $t$ (and therefore has better theoretical convergence properties). However, in our case increasing $t_0$ results in a very rapid growth of noise as a function of $t$ and obtaining reliable GEVP plateaux with this method would require a huge increase of statistics.

\subsubsection{Finding plateau ranges for the mass extraction}

The masses of non-zero electric flux states are extracted in the usual way by looking for plateaux in the effective masses obtained from Polyakov-loop correlators. The plateau range is fixed by first determining the value of $t$ where the effective mass is minimum among points with relative statistical error smaller than $4\%$. The fitting range around that time is then adjusted to keep a $\chi^2$ per degree of freedom smaller than 1. In most cases we take characteristic plateau ranges  of 5 to 8 points with resulting $\chi^2$ per degree of freedom below 0.5. The error is determined from the dispersion in fit parameters obtained by repeating this fitting procedure in jackknife bins.

The correlators used for the mass determination for vanishing electric flux are more noisy, also the energies are in some cases very closely spaced (level crossings), therefore a more systematic way of choosing the plateau ranges was introduced. We use single-exponential correlated fits, and the time ranges are selected using a variation of ``fit histogram'' method from Ref.~\cite{Ryan:2015gda}. For each $t_\mathrm{min}, t_\mathrm{max}$ we calculate the quantity measuring the ``fit quality'':
\begin{equation}
Q = \Gamma(N_\dof/2,\chi^2/2) N_\dof\;\! \tmin / \Delta E,
\end{equation}
where $\Gamma$ is the cumulative distribution function of the appropriate \mbox{$\chi^2$-distribution}. Next we choose the fit interval which maximizes $Q$. We also analyse whether the mass gap  fit result is stable among the fits with highest fit quality. In some cases, statistics had to be increased to fulfill that condition.

On the other hand, for the next excited state this condition cannot always be fulfilled without a significant increase of the statistics, therefore we calculate the standard deviation of ten fit results with the highest $Q$ and quote that as an estimate of the systematic error due to the choice of the fit range.

\subsubsection{Selection of glueball states}

As discussed in sec.~\ref{s.glueselect} we have to face the problem that the glueball state is not always
the lowest state over the vacuum in the zero electric flux sector due to the presence of light
torelon-pair states. The
identification of the glueball has required a procedure to determine the operator content of each state.
We expect that torelon-pairs couple mostly to Polyakov loop operators while the glueball state would
project more onto Wilson loop operators. In what follows we will briefly discuss how to implement this
selection procedure.

Suppose we have a collection of normalized states  $|\alpha \rangle$,
which have been identified as eigenstates of the Hamiltonian, and
a set of operators $\mathbf{O}_i$ that produce these states when acting on the vacuum. Now we can define a measure of the overlap
of a given operator with a given eigenstate of the Hamiltonian
as follows:
\be
\tilde{V}(\mathbf{O}_i, \alpha)= \frac{| \langle 0| \mathbf{O}_i | \alpha|
\rangle|}{\sqrt{\langle 0| \mathbf{O}_i \mathbf{O}_i^\dagger| 0 \rangle}}.
\label{eq:overlap}
\ee
The number in question is positive and smaller than 1. When this
number becomes large we can say that we have a large overlap between
the state (for example the lowest energy state) and the operator $\mathbf{O_i}$. 

The idea can be also applied in the case in which there are several
operators having similar properties, like for example using different
smearing levels. The philosophy is to gather all the operators of the
same type into a given set. For example, suppose we have a set $G$ as
follows
\be
G=\{\mathbf{O}_1, \ldots \mathbf{O}_p\}.
\ee
We can associate a vector space $V_G$ to the set $G$. This is given by
the $p$ dimensional subspace of the total Hilbert space generated by
the vectors $\mathbf{O}_i|0\rangle$. Now we can define a measure of the overlap
of these set of operators with a given eigenstate of the Hamiltonian
as follows:
\be
\tilde{V}(G,\alpha)= \max_{u\in V_G;\, ||u||=1} | (u ,|\alpha\rangle)|
\ee
where the maximized quantity is the absolute value of  the Hilbert space scalar product of the vector $u$
with the state $|\alpha\rangle$. Notice for a single operator it coincides with the previous formula.

The  two lowest lying eigenstates of the Hamiltonian and their
coupling to operators can be obtained from the results of the 
GEVP method described in sec.~\ref{s.determination}. This approach 
seems to work remarkably well when working with a reduced family of
operators including only one Wilson loop operator and one  operator
constructed from Polyakov loops for each torelon pair.

\subsubsection{Suppression of finite $T$ contributions in glueball correlation functions}
\label{s.finiteT}
One problem necessary to overcome when dealing with glueball correlation functions is the effect of the wrap-around states, coming from the finite time extent of the lattice. In perturbation theory one can show that the leading effect is a single wrap-around gluon, giving rise to a constant term in the correlators (even in the connected ones) proportional to $\exp(-m_t T)$, where $m_t$ is the mass of the lightest torelon state. While methods similar to the ones used in Ref.~\cite{Umeda:2007hy}, like midpoint subtraction of the correlators can be used to mitigate this problem, we find it most efficient to always use lattices with large enough $T$, that the contribution from those states is negligible.

\newpage

\FloatBarrier

\section{Tables: non-perturbative data}
\label{s.tables}

\subsection{Lattice data for the torelon spectrum}
\label{s.tablese}

\begin{table}[h]
\begin{center}
\begin{tabular}{cccccc}
$|\kbar|$  & $L$ & $b$ &$x$ & $E(1)$ & $E(2) $ \\
\hline
1 &14   & 28.00  & 0.19894  &  0.0870(22)  & 0.1774(73)  \\
   1 &14   & 18.00  & 0.30947  &  0.0861(19)  & 0.1730(66)  \\
   1 &14   & 10.00  & 0.55704  &  0.0854(19)  & 0.1746(55)  \\  
   1 &14   &  8.00  & 0.69630  &  0.0839( 9)  & 0.1695(43)  \\
   1 &14   &  6.00  & 0.92840  &  0.0787( 7)  & 0.1690(38)  \\
   1 &14   &  4.50  & 1.23787  &  0.0762(12)  & 0.1608(41)  \\
   1 &14   &  3.50  & 1.59155  &  0.0785( 8)  & 0.1608(41)  \\
\hline

\end{tabular}
\caption{$SU(5)$ torelon energies (in lattice units). The states are labeled by the value of the electric flux modulus $E (|\vec e|)$.}
\label{tab:n5_el}
\end{center}
\end{table}

\begin{table}[h]
\begin{center}
\begin{tabular}{ccccccc}
$|\kbar|$   & $L$ & $b$ &$x$ & $E(1)$ & $E(2)$  & $E(3)$ \\
\hline
1 &  10  &   28.00  &  0.19894  & 0.0851 (33) & 0.1736(52) & 0.2682(10) \\
   1 &  10  &   18.00  &  0.30947  & 0.0869 (17) & 0.1788(42) & 0.2667(13) \\
   1 &  10  &   10.00  &  0.55704  & 0.08466(75) & 0.1761(26) & 0.2622(57) \\
   1 &  10  &    8.00  &  0.69630  & 0.0808 (15) & 0.173 (16) & 0.2485(99) \\
   1 &  10  &    6.00  &  0.92840  & 0.07797(45) & 0.1663(69) & 0.2595(58) \\
   1 &  10  &    4.50  &  1.23787  & 0.0702 (10) & 0.1548(91) & 0.2479(94) \\
   1 &  10  &    3.50  &  1.59155  & 0.06664(34) & 0.1480(35) & 0.2244(91) \\
\end{tabular}
\caption{$SU(7)$ torelon energies (in lattice units). The states are labeled by the value of the electric flux modulus.}
\label{tab:n7_el}
\end{center}
\end{table}

\begin{table}[h]
\begin{center}
\begin{tabular}{ccccccc}
$|\kbar|$   & $L$ & $b$ & $x$ & $E(1)$ & $E(2)$  & $E(3)$ \\
\hline
3 &  10  &   28.00   & 0.19894  & 0.1683(49)  &  0.2666(45) &  0.0871(12) \\
   3 &  10  &   18.00   & 0.30947  & 0.1783(16)  &  0.2690(46) &  0.0888(15) \\
   3 &  10  &   10.00   & 0.55704  & 0.1741(95)  &  0.258 (18) &  0.0855(13) \\
   3 &  10  &    8.00   & 0.69630  & 0.1760(15)  &  0.2689( 5) &  0.0853( 9) \\
   3 &  10  &    6.00   & 0.92840  & 0.1738( 9)  &  0.2453(95) &  0.0824( 6) \\
   3 &  10  &    4.50   & 1.23787  & 0.1698(23)  &  0.2766(47) &  0.0888(27) \\
   3 &  10  &    3.50   & 1.59155  & 0.1674(50)  &  0.2724(62) &  0.1063(24) \\
   3 &  10  &    3.00   & 1.85681  & 0.1586(36)  &  0.2561(80) &  0.1309(44) \\
   3 &  10  &    2.50   & 2.22817  & 0.1604(53)  &  0.278 (10) &  0.1756(14) \\
   3 &  10  &    2.25   & 2.47574  & 0.1687(37)  &  0.2868(65) &  0.2045(60) \\
   3 &  10  &    2.00   & 2.78521  & 0.1673(61)  &  0.3148(28) &  0.2598(51) \\
   3 &  20  &    4.00   & 2.78521  & 0.0896(14)  &  0.1505(63) &  0.1249(38) \\
   3 &  10  &    1.75   & 3.18310  & 0.1938(14)  &  0.3520(48) &  0.3247(63) \\
   3 &  10  &    1.50   & 3.71362  & 0.2266(29)  &  0.4299(67) &  0.4377(55) \\
   3 &  20  &    2.50   & 4.45634  & 0.1356(54)  &  0.2682(62) &  0.294 (11) \\
   3 &  20  &    2.00   & 5.57042  & 0.2076(67)  &  0.4235(67) &  0.484 (48) \\
   3 &  20  &    1.75   & 6.36620  & 0.274 (28)  &  0.533 (29) &  0.680 (10) \\
\hline

\\
\end{tabular}
\caption{$SU(7)$ torelon energies (in lattice units) (continued). }
\label{tab:n7_el_b}
\end{center}
\end{table}

\begin{table}[h]
\begin{center}
\begin{tabular}{cccccccc}
$|\kbar|$  & $L$ & $x$ &  $E(1)$ & $E(2)$  & $E(3)$ & $E(4)$ & $E(5)$ \\
\hline
2 &  6  &  0.19894  & 0.458 (11) & 0.0926(13) & 0.3775(95) & 0.1893(26) & 0.2858(15) \\
2 &  6  &  0.55704  & 0.4778(55) & 0.0890(43) & 0.3756(50) & 0.1841(37) & 0.2824(80) \\
2 &  6  &  0.92840  & 0.468 (20) & 0.0829(12) & 0.3749(55) & 0.1790(44) & 0.2844(54) \\
2 &  6  &  1.23787  & 0.4566(63) & 0.0745(15) & 0.3715(37) & 0.1682(66) & 0.2641(95) \\
2 &  6  &  1.59155  & 0.4594(38) & 0.0803( 6) & 0.3672(70) & 0.1610(59) & 0.2671(47) \\
2 &  6  &  1.85681  & 0.4596(61) & 0.0843(86) & 0.350 (12) & 0.1750(40) & 0.243 (11) \\
2 &  6  &  2.22817  & 0.4467(93) & 0.1025( 9) & 0.3777(70) & 0.2028(40) & 0.2843(92) \\
\end{tabular}
\caption{$SU(11)$ torelon energies (in lattice units). The states are labelled by the value of the electric flux modulus.}
\label{tab:n11_el}
\end{center}
\end{table}

\begin{table}[h]
\begin{center}
\begin{tabular}{cccccccc}
$|\kbar|$ & $L$ &  $x$ &  $E(1)$ & $E(2)$  & $E(3)$ & $E(4)$ & $E(5)$ \\
\hline
4 &        6 &   0.19894  & 0.272 (12) & 0.4689( 9)  &  0.1841(61)  &  0.0930(19)  &  0.3768(40) \\
   4 &        6 &   0.55704  & 0.2774(57) & 0.467 (11)  &  0.1877(30)  &  0.0868(37)  &  0.373 (13) \\
   4 &        6 &   0.92840  &   -    -   & 0.4610(98)  &  0.1804(97)  &  0.0877(16)  &  0.368 (11) \\
   4 &        6 &   1.23787  & 0.272 (10) & 0.4694(34)  &  0.1813(27)  &  0.0893(26)  &  0.3734(69) \\
   4 &        6 &   1.59155  & 0.2718(20) & 0.4670(41)  &  0.1767(63)  &  0.1058(31)  &  0.3881(99) \\
   4 &        6 &   1.85681  & 0.2697(93) & 0.455 (10)  &  0.1919(17)  &  0.1249(52)  &  0.4108(41) \\
   4 &        6 &   2.22817  & 0.2649(66) & 0.567 (49)  &  0.2048(48)  &  0.1624(73)  &  0.4158(59) \\
\end{tabular}
\caption{$SU(11)$ torelon energies (in lattice units) (continued).}
\label{tab:n11_el_b}
\end{center}
\end{table}

\begin{table}[h]
\begin{center}
\begin{tabular}{ccccccccc}
$|\kbar|$ & $L$ &  $x$ & $E(1)$ & $E(2)$  & $E(3)$ & $E(4)$ & $E(5)$ & $E(6)$\\
\hline
5 &        6 &     0.19894  & 0.3961(75) & 0.2371(45)  &  0.1590(68)  &  0.4614(82)  &  0.0783(21)             & 0.3133(83) \\
   5 &        6 &     0.55704  & 0.3962(18) & 0.2375(72)  &  0.1603(10)  &  0.4785(54)  &  0.0780( 6)             & 0.3174(70) \\
   5 &        6 &     0.92840  & 0.383 (25) & 0.2247(65)  &  0.1502(40)  &  0.4646(81)  &  0.0742(14)             & 0.3158(87) \\
   5 &        6 &     1.23787  & 0.3828(61) & 0.2329(61)  &  0.1544(82)  &  0.471 (12)  &  0.0777(17)             & 0.318 (11) \\
   5 &        6 &     1.59155  & 0.3893(57) & 0.2286(68)  &  0.1579(65)  &  0.479 (11)  &  0.0910(24)             & 0.330 (11) \\
   5 &        6 &     1.85681  & 0.371 (11) & 0.2349(23)  &  0.1583(54)  &  0.481 (15)  &  0.1095(37)             & 0.3385(78) \\
   5 &        6 &     2.22817  & 0.3867(85) & 0.2356(58)  &  0.1567(56)  &  0.4809(79)  &  0.1445(35)             & 0.3617(44) \\
   5 &        6 &     2.47574  & 0.383 (21) & 0.236 (17)  &  0.1786(33)  &  0.4985(96)  &  0.1760(38)             & 0.3793(87) \\
   5 &        6 &     2.78521  & 0.3885(87) & 0.2262(77)  &  0.2001(11)  &  0.512 (13)  &  0.2089(80)             & 0.411 (30) \\
   5 &       12 &     2.78521  & 0.2037(79) & 0.1133(44)  &  0.0965(24)  &  0.2673(69)  &  0.0991(30)             & 0.2004(76) \\
   5 &        6 &     3.18310  & 0.389 (11) & 0.2510(30)  &  0.2327(80)  &  0.5361(35)  &  0.281 (15)             & 0.4519(62) \\
   5 &       12 &     4.45634  & 0.1952(38) & 0.1604(20)  &  0.1900(95)  &  0.3418(80)  &  0.2573(52)             & 0.329 (14) \\
   5 &       12 &     5.57042  & 0.2052(60) & 0.223 (30)  &  0.2979(50)  &  0.4545(43)  &  0.4315(82)             & 0.4981(65) \\
   5 &       12 &     6.36620  & 0.2168(50) & 0.2783(85)  &  0.4031(99)  &  0.537 (14)  &    -    -       &    -   -   \\
\hline	

\\
\end{tabular}
\caption{$SU(13)$ torelon energies (in lattice units). The states are labelled by the value of the electric flux modulus.}
\label{tab:n13_el}
\end{center}
\end{table}

\begin{table}[h]
\begin{center}
\begin{tabular}{ccccccc}
$|\kbar|$  & $L$ &  $x$ &$E(1)$ & $E(2)$  & $E(3)$ & $E(4)$ \\
\hline
2 &        4 &     0.19894  &   -     -  & 0.0899(10)  &  0.619 (11)  &  0.1756(61)  \\
   2 &        4 &     0.55704  &   -     -  & 0.0855( 3)  &  0.6316(60)  &  0.1752(50)  \\ 
   2 &        4 &     0.92840  &   -     -  & 0.0767(15)  &  0.6307(88)  &  0.1626(52)  \\
   2 &        4 &     1.23787  &   -     -  & 0.0690( 3)  &  0.622 (11)  &  0.1492(47)  \\
   2 &        4 &     1.59155  &   -     -  & 0.0608( 6)  &  0.6223(34)  &  0.1302(65)  \\
   2 &        4 &     1.85681  &   -     -  & 0.0601( 3)  &  0.6275(59)  &  0.1327(35)  \\
   2 &        4 &     2.22817  &   -     -  & 0.0631( 5)  &  0.619 (17)  &  0.1339(29)  \\
   2 &        4 &     2.47574  &   -     -  & 0.0692( 7)  &  0.6299(14)  &  0.1492(40)  \\
   2 &        4 &     2.78521  &   -     -  & 0.0794( 8)  &  0.624 (12)  &  0.1663(94)  \\
   2 &        8 &     2.78521  & 0.360 (15) & 0.0387( 4)  &  0.3150(85)  &  0.0852(49)  \\
   2 &        4 &     3.18310  &   -     -  & 0.0993( 9)  &  0.6423(91)  &  0.2100(25)  \\
   2 &        4 &     3.71362  &   -     -  & 0.1314(38)  &  0.638 (26)  &  0.2875(44)  \\
   2 &        8 &     4.45634  & 0.341 (23) & 0.0912(24)  &  0.3628(58)  &  0.1864(30)  \\
   2 &        8 &     5.57042  & 0.3594(36) & 0.1523(51)  &  0.4131(71)  &  0.3226(41)  \\
   2 &        8 &     6.36620  & 0.361 (12) & 0.2048(71)  &  0.462 (12)  &  0.4286(66)  \\
\hline

\\
\end{tabular}
\caption{$SU(17)$ torelon energies (in lattice units). The states are labelled by the value of the electric flux modulus.}
\label{tab:n17_el}
\end{center}
\end{table}

\begin{table}[h]
\begin{center}
\begin{tabular}{ccccccc}
$|\kbar|$  & $L$ &  $x$ &$E(1)$ & $E(2)$  & $E(3)$ & $E(4)$ \\
\hline
3 &        4 &     0.19894  & 0.540 (10) &   -    -    &  0.0906(10)  &  0.628 (30)  \\
   3 &        4 &     0.55704  &   -    -   & 0.4524(11)  &  0.0876( 6)  &  0.624 (13)  \\
   3 &        4 &     0.92840  & 0.5311(93) & 0.455 (10)  &  0.0810( 8)  &  0.621 (14)  \\
   3 &        4 &     1.23787  & 0.5354(22) & 0.441 (19)  &  0.0740( 9)  &  0.625 (14)  \\
   3 &        4 &     1.59155  & 0.550 (37) & 0.456 (12)  &  0.0764( 5)  &  0.622 (13)  \\
   3 &        4 &     1.85681  & 0.5305(65) & 0.4385(84)  &  0.0825( 7)  &  0.6383(67)  \\
   3 &        4 &     2.22817  & 0.5308(27) & 0.441 (18)  &  0.0970(20)  &  0.6455(56)  \\
\end{tabular}
\caption{$SU(17)$ torelon energies (in lattice units) (continued).}
\label{tab:n17_el_c}
\end{center}
\end{table}

\begin{table}[h]
\begin{center}
\begin{tabular}{ccccccc}
$|\kbar|$  & $L$ & $x$ &$E(1)$ & $E(2)$  & $E(3)$ & $E(4)$ \\
\hline
5 &        4   &  0.19894  & 0.638 (15) & 0.2718(32)  &  0.3647(78)  &  0.5478(69)  \\
   5 &        4   &  0.55704  & 0.605 (15) & 0.2735(19)  &  0.3530(64)  &  0.5402(77)  \\
   5 &        4   &  0.92840  &  -    -    & 0.2732(46)  &  0.3601(80)  &  0.542 (13)  \\
   5 &        4   &  1.23787  & 0.6237(29) & 0.2637(85)  &  0.3516(54)  &  0.5453(12)  \\
   5 &        4   &  1.59155  & 0.6171(86) & 0.264 (11)  &  0.3649( 8)  &  0.542 (11)  \\
   5 &        4   &  1.85681  & 0.6043(65) & 0.2670(32)  &  0.361 (17)  &  0.522 (13)  \\
   5 &        4   &  2.22817  & 0.6268(33) & 0.2667(27)  &  0.3703(80)  &  0.542 (21)  \\
\end{tabular}
\caption{$SU(17)$ torelon energies (in lattice units) (continued). }
\label{tab:n17_el_e}
\end{center}
\end{table}

\begin{table}[h]
\begin{center}
\begin{tabular}{ccccccc}
$|\kbar|$ & $L$ & $x$ &$E(1)$ & $E(2)$  & $E(3)$ & $E(4)$ \\
\hline
7 &        4 &   0.19894    & 0.4580(20) & 0.624 (10)  &  0.1820(70)  &  0.2744(38)  \\
   7 &        4 &   0.55704    & 0.4527(26) & 0.6343(54)  &  0.1824(41)  &  0.2687(52)  \\
   7 &        4 &   0.92840    & 0.449 (18) & 0.6113(21)  &  0.1696(30)  &  0.2595(62)  \\
   7 &        4 &   1.23787    & 0.4439(41) & 0.6332(67)  &  0.1786(16)  &  0.2661(72)  \\
   7 &        4 &   1.59155    & 0.4432(29) & 0.631 (20)  &  0.1755(25)  &  0.2720(68)  \\
   7 &        4 &   1.85681    & 0.4403(71) & 0.6304(68)  &  0.1760(29)  &  0.2668(87)  \\
   7 &        4 &   2.22817    & 0.434 (16) & 0.623 (10)  &  0.1806(19)  &  0.285 (15)  \\
\end{tabular}
\caption{$SU(17)$ torelon energies (in lattice units) (continued).}
\label{tab:n17_el_e2}
\end{center}
\end{table}

\begin{table}[h]
\begin{center}
\begin{tabular}{ccccccc}
$|\kbar|$  & $L$ & $x$ & $E(5)$ & $E(6)$ & $E(7)$ & $E(8)$ \\
\hline
2 &        4   &  0.19894  & 0.5329(51) & 0.2711(57)  &  0.4526(46)  &  0.352 (10)  \\
   2 &        4   &  0.55704  & 0.5485(42) & 0.2676(49)  &  0.4475(64)  &  0.3584(39)  \\
   2 &        4   &  0.92840  & 0.5433(81) & 0.2679(42)  &  0.4557(19)  &  0.356 (11)  \\
   2 &        4   &  1.23787  & 0.536 (11) & 0.2478(50)  &  0.432 (15)  &  0.3489(94)  \\
   2 &        4   &  1.59155  & 0.5244(98) & 0.229 (22)  &  0.431 (15)  &  0.3356(28)  \\
   2 &        4   &  1.85681  & 0.514 (11) & 0.213 (13)  &  0.4023(80)  &  0.297 (11)  \\
   2 &        4   &  2.22817  & 0.5117(95) & 0.2350(45)  &  0.4281(66)  &  0.312 (13)  \\
\end{tabular}
\caption{$SU(17)$ torelon energies (in lattice units) (continued).}
\label{tab:n17_el_b}
\end{center}
\end{table}

\begin{table}[h]
\begin{center}
\begin{tabular}{ccccccc}
$|\kbar|$& $L$ &  $x$ & $E(5)$ & $E(6)$ & $E(7)$ & $E(8)$ \\
\hline
4 &        4   &  0.19894  & 0.2762(18) &   -     -   &  0.537 (13)  &  0.1850(87)  \\
   4 &        4   &  0.55704  & 0.2734(24) & 0.6372(13)  &  0.536 (14)  &  0.1856(31)  \\
   4 &        4   &  0.92840  & 0.2748(65) & 0.6342(54)  &  0.523 (12)  &  0.1742(47)  \\
   4 &        4   &  1.23787  & 0.2738(48) & 0.628 (40)  &  0.544 (12)  &  0.1724(17)  \\
   4 &        4   &  1.59155  & 0.2611(84) & 0.603 (12)  &  0.5459(66)  &  0.1737(46)  \\
   4 &        4   &  1.85681  & 0.2706(31) & 0.643 (16)  &  0.5694(50)  &  0.1818(53)  \\
   4 &        4   &  2.22817  & 0.2745(79) & 0.644 (11)  &  0.580 (20)  &  0.2120(21)  \\
\end{tabular}
\caption{$SU(17)$ torelon energies (in lattice units) (continued).}
\label{tab:n17_el_d}
\end{center}
\end{table}

\begin{table}[h]
\begin{center}
\begin{tabular}{ccccccc}
$|\kbar|$  & $L$ &  $x$ & $E(5)$ & $E(6)$ & $E(7)$ & $E(8)$ \\
\hline
6 &        4  &  0.19894  & 0.1860(80) & 0.0906(18)  &  0.3636(56)  &  0.6300(49)  \\
   6 &        4  &  0.55704  & 0.1817(28) & 0.0887(13)  &  0.3628(30)  &  0.6418(41)  \\
   6 &        4  &  0.92840  & 0.1770(50) & 0.0835(10)  &  0.3624(94)  &  0.6316(77)  \\
   6 &        4  &  1.23787  & 0.1803(43) & 0.0872(11)  &  0.3710(80)  &  0.637 (10)  \\
   6 &        4  &  1.59155  & 0.1792(53) & 0.1032(22)  &  0.3720(94)  &  0.6327(90)  \\
   6 &        4  &  1.85681  & 0.1825(67) & 0.1239(37)  &  0.3860(84)  &  0.6514(75)  \\
   6 &        4  &  2.22817  & 0.2059(39) & 0.160 (10)  &  0.410 (11)  &  0.690 (11)  \\
\end{tabular}
\caption{$SU(17)$ torelon energies (in lattice units) (continued).}
\label{tab:n17_el_f}
\end{center}
\end{table}

\begin{table}[h]
\begin{center}
\begin{tabular}{ccccccc}
$|\kbar|$ & $L$ & $x$ & $E(5)$ & $E(6)$ & $E(7)$ & $E(8)$ \\
\hline
8 &        4  &  0.19894  & 0.629 (11) & 0.438 (11)  &  0.2715(80)  &  0.0908( 6)  \\
   8 &        4  &  0.55704  & 0.624 (10) & 0.436 (13)  &  0.2712(37)  &  0.0874(18)  \\
   8 &        4  &  0.92840  & 0.6344(66) & 0.4408(72)  &  0.2617(58)  &  0.0816(12)  \\
   8 &        4  &  1.23787  &   -    -   & 0.4542(93)  &  0.2703(98)  &  0.0910(18)  \\
   8 &        4  &  1.59155  & 0.6700(92) & 0.458 (14)  &  0.2838(60)  &  0.1155(19)  \\
   8 &        4  &  1.85681  & 0.6361(45) & 0.4661(50)  &  0.2905(48)  &  0.1326(48)  \\
   8 &        4  &  2.22817  & 0.628 (12) & 0.4763(63)  &  0.315 (15)  &  0.1797(51)  \\
   8 &        4  &  2.47574  & 0.6438(80) & 0.476 (10)  &  0.3186(64)  &  0.2290(47)  \\
   8 &        4  &  2.78521  & 0.634 (14) & 0.469 (11)  &  0.3622(71)  &  0.283 (12)  \\
   8 &        8  &  2.78521  & 0.3315(89) & 0.2571(66)  &  0.1836(69)  &  0.1318(41)  \\
   8 &        4  &  3.18310  & 0.656 (24) & 0.505 (11)  &  0.399 (10)  &  0.3760(44)  \\
   8 &        4  &  3.71362  & 0.686 (14) & 0.584 (12)  &  0.541 (17)  &  0.5004(84)  \\
   8 &        8  &  4.45634  & 0.4084(81) & 0.3755(52)  &  0.3581(73)  &  0.368 (10)  \\
   8 &        8  &  5.57042  & 0.515 (42) & 0.510 (37)  &  0.5236(95)  &  0.6042(54)  \\
   8 &        8  &  6.36620  & 0.627 (33) & 0.6518(98)  &   -  -       &   -  -       \\
\hline

\\
\end{tabular}
\caption{$SU(17)$ torelon energies (in lattice units) (continued).}
\label{tab:n17_el_f2}
\end{center}
\end{table}

\begin{table}[h]
\begin{center}
\begin{tabular}{ccccc}
$|\kbar|$ & $L$ &  $x$ & $E(\sqrt 2)$ & $E(4\sqrt 2)$  \\
\hline
4 &        4   &  0.19894  & 0.5136(41)&0.1299(26)    \\
   4 &        4   &  0.55704  & 0.504 (12)&0.1257(33)    \\
   4 &        4   &  0.92840  & 0.484 (13)&0.1148(37)    \\
   4 &        4   &  1.23787  & 0.5118(45)&0.1209(28)    \\
   4 &        4   &  1.59155  & 0.515 (12)&0.1339(61)    \\
   4 &        4   &  1.85681  & 0.4849(94)&0.1579(23)    \\
   4 &        4   &  2.22817  & 0.491 (13)&0.1837(93)    \\
   4 &        4   &  2.47574  & 0.5002(65)&0.2210(45)    \\
   4 &        4   &  2.78521  & 0.5045(68)&0.2676(40)    \\
   4 &        8   &  2.78521  & 0.2449(64)&0.1249(37)    \\
   4 &        4   &  3.18310  & 0.4905(84)&0.3372(54)    \\
   4 &        8   &  4.45634  & 0.264 (78)&0.308 (17)    \\
   4 &        8   &  5.57042  & 0.2700(42)&0.5282(79)    \\
   4 &        8   &  6.36620  & 0.281 (13)& -  -     \\
\hline

\\
\end{tabular}
\caption{$SU(17)$ torelon energies (in lattice units) (continued).}
\label{tab:n17_el1}
\end{center}
\end{table}

\FloatBarrier

\subsection{Lattice data for the scalar glueball spectrum}
\label{s.tablesg}
\FloatBarrier

\begin{table}[h]
\begin{center}
\begin{tabular}{ccccc}
$|\kbar|$ & $L$ &  $x$ & $m_1$ & $m_2$\\
\hline
1 & 14 &  0.19894 & 0.1753(14) & 0.2572(38)(06) \\
1 & 14 &  0.30947 & 0.1738(12) & 0.2520(41)(02) \\
1 & 14 &  0.55704 & 0.16813(98) & 0.2496(28)(01) \\
1 & 14 &  0.69630 & 0.1593(31) & 0.2498(16)(02) \\
1 & 14 &  0.92840 & 0.1479(34) & 0.2530(28)(11) \\
1 & 14 &  1.23787 & 0.1470(21) & 0.2539(39)(77) \\
1 & 14 &  1.59155 & 0.1679(13) & 0.2700(24)(02) \\
1 & 14 &  1.85681 & 0.1895(15) & 0.2950(55)(04) \\
1 & 14 &  2.22817 & 0.2324(28) & 0.3618(43)(08) \\
1 & 14 &  2.47574 & 0.2693(10) & 0.4010(29)(35) \\
1 & 14 &  2.78521 & 0.3190(34) & 0.4449(71)(17) \\
1 & 28 &  2.78521 & 0.1540(27) & 0.2219(19)(12) \\
1 & 14 &  3.18310 & 0.4089(49) & 0.543(07)(17) \\
1 & 28 &  4.45634 & 0.3409(17) & 0.3818(27)(30) \\
1 & 28 &  5.57042 & 0.4286(68) & 0.620(12)(03) \\
1 & 28 &  6.36620 & 0.4905(83) & 0.7574(81)(24) \\
\hline
2 & 14 &  0.19894 & 0.1733(19) & 0.2522(21)(03) \\
2 & 14 &  0.30947 & 0.1714(11) & 0.25517(98)(24) \\
2 & 14 &  0.55704 & 0.1578(21) & 0.2570(18)(03) \\
2 & 14 &  0.69630 & 0.1463(16) & 0.2670(45)(09) \\
2 & 14 &  0.92840 & 0.1440(13) & 0.2787(13)(05) \\
2 & 14 &  1.23787 & 0.1693(24) & 0.2984(10)(03) \\
2 & 14 &  1.59155 & 0.2333(20) & 0.3121(61)(09) \\
2 & 14 &  1.85681 & 0.2909(17) & 0.3339(53)(05) \\
2 & 14 &  2.22817 & 0.3377(21) & 0.4138(25)(04) \\
2 & 14 &  2.47574 & 0.3641(18) & 0.416(21)(07) \\
2 & 14 &  2.78521 & 0.4083(48) & 0.5129(50)(06) \\
2 & 28 &  2.78521 & 0.2047(15) & 0.2586(22)(17) \\
2 & 14 &  3.18310 & 0.439(12) & 0.583(03)(21) \\
2 & 28 &  4.45634 & 0.3351(46) & 0.4030(68)(66) \\
2 & 28 &  5.57042 & 0.4262(38) & 0.646(04)(35) \\
2 & 28 &  6.36620 & 0.4948(37) & 0.761(08)(11) \\
\hline

\\
\end{tabular}
\caption{Mass gap and next excited state energy (in lattice units) in the zero-electric flux scalar sector for $SU(5)$.}
\label{tab:n5_glue}
\end{center}
\end{table}

\begin{table}[h]
\begin{center}
\begin{tabular}{ccccc}
$|\kbar|$ & $L$ &  $x$ & $m_1$  & $m_2$\\
\hline
1 & 10 &  0.19894 & 0.17576(80) & 0.2532(28)(08) \\
1 & 10 &  0.30947 & 0.1741(13) & 0.2444(45)(11) \\
1 & 10 &  0.55704 & 0.1646(17) & 0.2544(20)(09) \\
1 & 10 &  0.69630 & 0.16244(91) & 0.2545(23)(20) \\
1 & 10 &  0.92840 & 0.1494(40) & 0.2483(44)(10) \\
1 & 10 &  1.23787 & 0.13655(98) & 0.2498(64)(07) \\
1 & 10 &  1.59155 & 0.1362(11) & 0.247(12)(03) \\
1 & 10 &  1.85681 & 0.14832(68) & 0.3294(38)(73) \\
1 & 10 &  2.22817 & 0.1742(14) & 0.3274(92)(27) \\
1 & 10 &  2.47574 & 0.1939(12) & 0.375(03)(11) \\
1 & 10 &  2.78521 & 0.22401(90) & 0.393(12)(06) \\
1 & 20 &  2.78521 & 0.1067(25) & 0.2254(19)(03) \\
1 & 10 &  3.18310 & 0.2772(10) & 0.4889(90)(78) \\
1 & 20 &  4.45634 & 0.2433(13) & 0.3432(61)(12) \\
1 & 20 &  5.57042 & 0.3919(65) & 0.4370(37)(03) \\
1 & 20 &  6.36620 & 0.4927(38) & 0.5575(33)(90) \\
\hline
2 & 10 &  0.19894 & 0.1771(10) & 0.2521(20)(01) \\
2 & 10 &  0.30947 & 0.1723(16) & 0.2524(23)(01) \\
2 & 10 &  0.55704 & 0.1635(24) & 0.2554(15)(01) \\
2 & 10 &  0.69630 & 0.1542(15) & 0.2536(20)(01) \\
2 & 10 &  0.92840 & 0.1455(13) & 0.2415(53)(13) \\
2 & 10 &  1.23787 & 0.1620(11) & 0.2596(30)(04) \\
2 & 10 &  1.59155 & 0.2032(22) & 0.2814(23)(07) \\
2 & 10 &  1.85681 & 0.2429(36) & 0.3189(66)(11) \\
2 & 10 &  2.22817 & 0.3154(21) & 0.3689(25)(12) \\
2 & 10 &  2.47574 & 0.3698(28) & 0.3986(40)(09) \\
2 & 10 &  2.78521 & 0.4493(16) & 0.4398(90)(15) \\
2 & 20 &  2.78521 & 0.2173(34) & 0.2245(14)(09) \\
2 & 10 &  3.18310 & 0.4825(87) & 0.5634(25)(26) \\
2 & 20 &  4.45634 & 0.3347(26) & 0.3907(48)(72) \\
2 & 20 &  5.57042 & 0.4353(71) & 0.5001(45)(07) \\
2 & 20 &  6.36620 & 0.5030(63) & 0.578(25)(04) \\
\hline

\\
\end{tabular}
\caption{Mass gap and next excited state energy (in lattice units) in the zero-electric flux scalar sector for $SU(7)$.}
\label{tab:n7_glue}
\end{center}
\end{table}

\begin{table}[h]
\begin{center}
\begin{tabular}{ccccc}
$|\kbar|$ & $L$ &  $x$ & $m_1$  & $m_2$\\
\hline
3 & 10 &  0.19894 & 0.1724(13) & 0.2521(26)(01) \\
3 & 10 &  0.30947 & 0.1713(13) & 0.2561(14)(01) \\
3 & 10 &  0.55704 & 0.1569(10) & 0.2585(26)(03) \\
3 & 10 &  0.69630 & 0.1456(42) & 0.2624(48)(08) \\
3 & 10 &  0.92840 & 0.14359(95) & 0.2820(25)(04) \\
3 & 10 &  1.23787 & 0.1694(30) & 0.3073(44)(08) \\
3 & 10 &  1.59155 & 0.2455(26) & 0.3211(24)(06) \\
3 & 10 &  1.85681 & 0.289(12) & 0.3316(18)(02) \\
3 & 10 &  2.22817 & 0.3243(26) & 0.3839(86)(09) \\
3 & 10 &  2.47574 & 0.3316(27) & 0.392(14)(01) \\
3 & 10 &  2.78521 & 0.3603(29) & 0.4629(22)(05) \\
3 & 20 &  2.78521 & 0.1778(32) & 0.2344(16)(05) \\
3 & 10 &  3.18310 & 0.3963(39) & 0.521(10)(01) \\
3 & 20 &  4.45634 & 0.3002(20) & 0.3523(15)(04) \\
3 & 20 &  5.57042 & 0.4386(39) & 0.4523(40)(10) \\
3 & 20 &  6.36620 & 0.4782(87) & 0.5995(60)(10) \\
\hline

\\
\end{tabular}
\caption{Mass gap and next excited state energy (in lattice units) in the zero-electric flux scalar sector for $SU(7)$ (continued).}
\label{tab:n7b_glue}
\end{center}
\end{table}

\begin{table}[h]
\begin{center}
\begin{tabular}{ccccc}
$|\kbar|$ &  $L$ & $x$ & $m_1$  & $m_2$\\
\hline
1  & 6 & 0.19894 & 0.1837(10) & 0.2673(19)(01) \\
1  & 6 & 0.55704 & 0.1718(12) & 0.2695(40)(02) \\
1  & 6 & 0.92840 & 0.1516(19) & 0.2666(39)(03) \\
1  & 6 & 1.23787 & 0.1259(14) & 0.2656(24)(02) \\
1  & 6 & 1.59155 & 0.0969(12) & 0.2413(48)(33) \\
1  & 6 & 1.85681 & 0.08532(75) & 0.229(12)(01) \\
1  & 6 & 2.22817 & 0.08615(77) & 0.2788(67)(17) \\
1  & 6 & 2.47574 & 0.0949(14) & 0.3097(57)(15) \\
1  & 6 & 2.78521 & 0.1114(11) & 0.3516(48)(39) \\
1  & 12 & 2.78521 & 0.0547(15) & 0.2141(39)(06) \\
1  & 6 & 3.18310 & 0.13907(79) & 0.4115(68)(11) \\
1  & 12 & 4.45634 & 0.1325(14) & 0.3717(43)(06) \\
1  & 12 & 5.57042 & 0.2199(47) & 0.4759(52)(09) \\
1  & 12 & 6.36620 & 0.3159(14) & 0.534(03)(38) \\
\hline

\\
\end{tabular}
\caption{Mass gap and next excited state energy (in lattice units) in the zero-electric flux scalar sector for $SU(11)$.}
\label{tab:n11_glue}
\end{center}
\end{table}

\begin{table}[h]
\begin{center}
\begin{tabular}{ccccc}
$|\kbar|$ &  $L$ & $x$ & $m_1$  & $m_2$\\
\hline
2  & 6 & 0.19894 & 0.1861(14) & 0.2702(32)(03) \\
2  & 6 & 0.55704 & 0.1766(17) & 0.2698(15)(10) \\
2  & 6 & 0.92840 & 0.1546(21) & 0.2683(31)(05) \\
2  & 6 & 1.23787 & 0.1497(20) & 0.2616(40)(02) \\
2  & 6 & 1.59155 & 0.1685(14) & 0.2830(19)(11) \\
2  & 6 & 1.85681 & 0.1906(11) & 0.3059(42)(03) \\
2  & 6 & 2.22817 & 0.23004(87) & 0.3521(70)(28) \\
2  & 6 & 2.47574 & 0.2605(19) & 0.4198(59)(09) \\
2  & 6 & 2.78521 & 0.3144(13) & 0.4816(17)(01) \\
2  & 12 & 2.78521 & 0.1454(54) & 0.2361(33)(06) \\
2  & 6 & 3.18310 & 0.3873(31) & 0.5573(69)(09) \\
2  & 12 & 4.45634 & 0.3376(39) & 0.3776(19)(59) \\
2  & 12 & 5.57042 & 0.4431(94) & 0.5485(36)(13) \\
2  & 12 & 6.36620 & 0.507(11) & 0.579(09)(15) \\
\hline
3  & 6 & 0.19894 & 0.18737(70) & 0.2627(24)(07) \\
3  & 6 & 0.55704 & 0.1755(18) & 0.2681(38)(05) \\
3  & 6 & 0.92840 & 0.1538(14) & 0.2648(16)(04) \\
3  & 6 & 1.23787 & 0.1728(13) & 0.2675(35)(07) \\
3  & 6 & 1.59155 & 0.2121(15) & 0.3018(12)(04) \\
3  & 6 & 1.85681 & 0.2490(29) & 0.3333(23)(09) \\
3  & 6 & 2.22817 & 0.3241(15) & 0.3951(21)(04) \\
3  & 6 & 2.47574 & 0.3783(31) & 0.4322(19)(06) \\
3  & 6 & 2.78521 & 0.4547(30) & 0.4789(35)(05) \\
3  & 12 & 2.78521 & 0.2230(40) & 0.2393(22)(02) \\
3  & 6 & 3.18310 & 0.5267(37) & 0.5827(30)(02) \\
3  & 12 & 4.45634 & 0.3555(37) & 0.3953(31)(77) \\
3  & 12 & 5.57042 & 0.4168(64) & 0.486(10)(02) \\
3  & 12 & 6.36620 & 0.4894(37) & 0.5462(26)(08) \\
\hline

\\
\end{tabular}
\caption{Mass gap and next excited state energy (in lattice units) in the zero-electric flux scalar sector for $SU(11)$ (continued).}
\label{tab:n11a_glue}
\end{center}
\end{table}

\begin{table}[h]
\begin{center}
\begin{tabular}{ccccc}
$|\kbar|$ &  $L$ & $x$ & $m_1$  & $m_2$\\
\hline
4 &  6 & 0.19894 & 0.1846(15) & 0.2687(19)(01) \\
4 &  6 & 0.30947 & 0.1830(14) & 0.2644(34)(14) \\
4 &  6 & 0.55704 & 0.1708(21) & 0.2790(12)(11) \\
4 &  6 & 0.69630 & 0.1585(19) & 0.2812(25)(02) \\
4 &  6 & 0.92840 & 0.1528(26) & 0.2874(26)(02) \\
4 &  6 & 1.23787 & 0.1837(17) & 0.3046(12)(01) \\
4 &  6 & 1.59155 & 0.2438(22) & 0.3179(47)(03) \\
4 &  6 & 1.85681 & 0.2942(17) & 0.3411(36)(32) \\
4 &  6 & 2.22817 & 0.3673(21) & 0.4131(20)(04) \\
4 &  6 & 2.47574 & 0.4114(24) & 0.4774(19)(08) \\
4 &  6 & 2.78521 & 0.4604(37) & 0.5325(31)(03) \\
4 &  12 & 2.78521 & 0.2327(28) & 0.2694(18)(01) \\
4 &  6 & 3.18310 & 0.4971(64) & 0.593(14)(05) \\
4 &  12 & 4.45634 & 0.2895(22) & 0.3757(13)(03) \\
4 &  12 & 5.57042 & 0.359(10) & 0.4616(81)(14) \\
4 &  12 & 6.36620 & 0.4239(57) & 0.534(04)(14) \\
\hline
5 &  6 & 0.19894 & 0.1844(12) & 0.2687(16)(01) \\
5 &  6 & 0.55704 & 0.1640(31) & 0.2863(21)(13) \\
5 &  6 & 0.92840 & 0.1556(13) & 0.3067(30)(08) \\
5 &  6 & 1.23787 & 0.1886(15) & 0.3313(39)(06) \\
5 &  6 & 1.59155 & 0.2571(20) & 0.3453(28)(20) \\
5 &  6 & 1.85681 & 0.3194(23) & 0.3374(12)(01) \\
5 &  6 & 2.22817 & 0.3229(50) & 0.3960(34)(09) \\
5 &  6 & 2.47574 & 0.3120(29) & 0.4282(23)(01) \\
5 &  6 & 2.78521 & 0.3137(20) & 0.4732(62)(18) \\
5 &  12 & 2.78521 & 0.1590(18) & 0.2347(49)(03) \\
5 &  6 & 3.18310 & 0.3176(45) & 0.5292(51)(12) \\
5 &  12 & 3.71362 & 0.17637(74) & 0.3047(19)(15) \\
5 &  12 & 4.45634 & 0.2073(18) & 0.3626(24)(13) \\
5 &  12 & 5.57042 & 0.2852(36) & 0.4721(37)(34) \\
5 &  12 & 6.36620 & 0.3622(20) & 0.5292(51)(12) \\
\hline

\\
\end{tabular}
\caption{Mass gap and next excited state energy (in lattice units) in the zero-electric flux scalar sector for $SU(11)$ (continued).}
\label{tab:n11b_glue}
\end{center}
\end{table}

\begin{table}[h]
\begin{center}
\begin{tabular}{ccccc}
$|\kbar|$ & $L$ & $x$ & $m_1$  & $m_2$\\
\hline
5 &  6 & 0.19894 & 0.1574(11) & 0.2259(16)(01) \\
5 &  6 & 0.55704 & 0.1415(16) & 0.2301(12)(01) \\
5 &  6 & 0.92840 & 0.1318(18) & 0.2376(22)(06) \\
5 &  6 & 1.23787 & 0.1583(24) & 0.2604(21)(07) \\
5 &  6 & 1.59155 & 0.2138(14) & 0.2794(29)(01) \\
5 &  6 & 1.85681 & 0.2592(19) & 0.2962(56)(04) \\
5 &  6 & 2.22817 & 0.3003(34) & 0.3620(26)(46) \\
5 &  6 & 2.47574 & 0.3392(20) & 0.3980(71)(05) \\
5 &  6 & 2.78521 & 0.3845(32) & 0.4195(89)(30) \\
5 &  12 & 2.78521 & 0.1957(31) & 0.2315(32)(06) \\
5 &  6 & 3.18310 & 0.4405(63) & 0.5102(29)(82) \\
5 &  12 & 4.45634 & 0.3058(49) & 0.3392(51)(25) \\
5 &  12 & 5.57042 & 0.3883(30) & 0.4297(25)(48) \\
5 &  12 & 6.36620 & 0.4231(56) & 0.4732(51)(24) \\
\hline

\\
\end{tabular}
\caption{Mass gap and next excited state energy (in lattice units) in the zero-electric flux scalar sector for $SU(13)$.}
\label{tab:n13_glue}
\end{center}
\end{table}

\begin{table}[h]
\begin{center}
\begin{tabular}{ccccc}
$|\kbar|$ &  $L$ & $x$ & $m_1$  & $m_2$\\
\hline
2 &  4 & 0.19894 & 0.17974(99) & 0.2579(17)(21) \\
2 &  4 & 0.55704 & 0.1700(16) & 0.2601(25)(10) \\
2 &  4 & 0.92840 & 0.1507(10) & 0.2614(35)(04) \\
2 &  4 & 1.23787 & 0.1311(13) & 0.2510(55)(03) \\
2 &  4 & 1.59155 & 0.1223(12) & 0.2597(49)(07) \\
2 &  4 & 1.85681 & 0.1259(11) & 0.2771(37)(59) \\
2 &  4 & 2.22817 & 0.1397(10) & 0.3306(23)(98) \\
2 &  4 & 2.47574 & 0.1566(17) & 0.357(05)(13) \\
2 &  4 & 2.78521 & 0.1773(26) & 0.4052(52)(08) \\
2 &  8 & 2.78521 & 0.08683(80) & 0.2069(49)(10) \\
2 &  4 & 3.18310 & 0.2192(18) & 0.5086(35)(07) \\
2 &  8 & 4.45634 & 0.1951(18) & 0.3701(19)(15) \\
2 &  8 & 5.57042 & 0.3246(35) & 0.455(02)(17) \\
2 &  8 & 6.36620 & 0.4336(68) & 0.5194(40)(46) \\
\hline

\\
\end{tabular}
\caption{Mass gap and next excited state energy (in lattice units) in the zero-electric flux scalar sector for $SU(17)$.}
\label{tab:n17_glue}
\end{center}
\end{table}
\begin{table}[h]
\begin{center}
\begin{tabular}{ccccc}
$|\kbar|$ &  $L$ & $x$ & $m_1$  & $m_2$\\
\hline
3 &  4 & 0.19894 & 0.1822(12) & 0.2604(23)(22) \\
3 &  4 & 0.55704 & 0.1716(22) & 0.2616(30)(05) \\
3 &  4 & 0.92840 & 0.1538(17) & 0.2693(30)(48) \\
3 &  4 & 1.23787 & 0.1456(13) & 0.2486(37)(08) \\
3 &  4 & 1.59155 & 0.1605(12) & 0.2611(73)(12) \\
3 &  4 & 1.85681 & 0.18125(92) & 0.3001(33)(13) \\
3 &  4 & 2.22817 & 0.2171(17) & 0.3602(27)(07) \\
3 &  4 & 2.47574 & 0.2481(11) & 0.4084(27)(08) \\
3 &  4 & 2.78521 & 0.2897(30) & 0.4580(38)(02) \\
3 &  8 & 2.78521 & 0.1370(26) & 0.2384(23)(04) \\
3 &  4 & 3.18310 & 0.3635(36) & 0.5354(44)(07) \\
3 &  8 & 4.45634 & 0.3223(62) & 0.3527(90)(24) \\
3 &  8 & 5.57042 & 0.4388(48) & 0.5306(26)(84) \\
3 &  8 & 6.36620 & 0.4980(71) & 0.5548(28)(05) \\
\hline
4 &  4 & 0.19894 & 0.1810(19) & 0.2594(16)(09) \\
4 &  4 & 0.55704 & 0.1719(14) & 0.2590(16)(10) \\
4 &  4 & 0.92840 & 0.1494(21) & 0.2549(17)(23) \\
4 &  4 & 1.23787 & 0.1594(15) & 0.2548(31)(13) \\
4 &  4 & 1.59155 & 0.1891(16) & 0.2659(41)(19) \\
4 &  4 & 1.85681 & 0.2213(16) & 0.3155(52)(03) \\
4 &  4 & 2.22817 & 0.2782(20) & 0.3732(29)(04) \\
4 &  4 & 2.47574 & 0.3185(32) & 0.4128(60)(09) \\
4 &  4 & 2.78521 & 0.3863(22) & 0.4670(23)(23) \\
4 &  8 & 2.78521 & 0.1832(31) & 0.2240(67)(07) \\
4 &  4 & 3.18310 & 0.4887(66) & 0.513(13)(05) \\
4 &  8 & 4.45634 & 0.3342(35) & 0.3614(26)(05) \\
4 &  8 & 5.57042 & 0.3427(36) & 0.4585(44)(07) \\
4 &  8 & 6.36620 & 0.3681(49) & 0.5196(62)(22) \\
\hline
\\
\end{tabular}
\caption{Mass gap and next excited state energy (in lattice units) in the zero-electric flux scalar sector for $SU(17)$ (continued).}
\label{tab:n17a_glue}
\end{center}
\end{table}

\begin{table}[h]
\begin{center}
\begin{tabular}{ccccc}
$|\kbar|$ &  $L$ & $x$ & $m_1$  & $m_2$\\
\hline
5 &  4 & 0.19894 & 0.1808(16) & 0.2592(16)(14) \\
5 &  4 & 0.55704 & 0.16898(78) & 0.2632(19)(40) \\
5 &  4 & 0.92840 & 0.1486(19) & 0.2632(14)(02) \\
5 &  4 & 1.23787 & 0.1637(22) & 0.2702(23)(05) \\
5 &  4 & 1.59155 & 0.2167(14) & 0.2887(44)(10) \\
5 &  4 & 1.85681 & 0.2585(17) & 0.3276(30)(04) \\
5 &  4 & 2.22817 & 0.3315(15) & 0.3847(11)(16) \\
5 &  4 & 2.47574 & 0.3938(26) & 0.418(03)(15) \\
5 &  4 & 2.78521 & 0.4596(28) & 0.4791(39)(09) \\
5 &  8 & 2.78521 & 0.2258(75) & 0.2335(34)(01) \\
5 &  4 & 3.18310 & 0.4823(79) & 0.5653(36)(03) \\
5 &  8 & 4.45634 & 0.3174(23) & 0.3651(61)(07) \\
5 &  8 & 5.57042 & 0.4586(46) & 0.6288(70)(06) \\
5 &  8 & 6.36620 & 0.5196(35) & 0.514(07)(16) \\
\hline
6 &  4 & 0.19894 & 0.1775(23) & 0.2613(18)(05) \\
6 &  4 & 0.30947 & 0.1765(17) & 0.2580(21)(04) \\
6 &  4 & 0.55704 & 0.1687(26) & 0.2661(34)(01) \\
6 &  4 & 0.69630 & 0.1539(19) & 0.2660(17)(02) \\
6 &  4 & 0.92840 & 0.1477(11) & 0.2660(36)(12) \\
6 &  4 & 1.23787 & 0.1780(16) & 0.2800(33)(17) \\
6 &  4 & 1.59155 & 0.2306(16) & 0.3066(27)(08) \\
6 &  4 & 1.85681 & 0.2782(28) & 0.3324(24)(03) \\
6 &  4 & 2.22817 & 0.3468(36) & 0.3927(21)(05) \\
6 &  4 & 2.47574 & 0.3995(20) & 0.4527(20)(17) \\
6 &  4 & 2.78521 & 0.4525(32) & 0.4896(21)(03) \\
6 &  8 & 2.78521 & 0.2304(45) & 0.2466(20)(03) \\
6 &  4 & 3.18310 & 0.4722(69) & 0.521(03)(16) \\
6 &  8 & 3.71362 & 0.23836(94) & 0.3008(32)(13) \\
6 &  8 & 4.45634 & 0.2314(22) & 0.3505(74)(05) \\
6 &  8 & 5.57042 & 0.2453(59) & 0.4587(59)(08) \\
6 & 16 & 5.57042 & 0.1240(24) & 0.2320(28)(28) \\
6 &  8 & 6.36620 & 0.2695(21) & 0.5030(64)(22) \\
6 & 16 & 11.14085 & 0.297(18) & 0.550(38)(09) \\
\hline

\\
\end{tabular}
\caption{Mass gap and next excited state energy (in lattice units) in the zero-electric flux scalar sector for $SU(17)$ (continued).}
\label{tab:n17b_glue}
\end{center}
\end{table}

\begin{table}[h]
\begin{center}
\begin{tabular}{ccccc}
$|\kbar|$ &  $L$ & $x$ & $m_1$  & $m_2$\\
\hline
7 &  4 & 0.19894 & 0.1796(13) & 0.2618(12)(02) \\
7 &  4 & 0.30947 & 0.1745(17) & 0.2625(17)(33) \\
7 &  4 & 0.55704 & 0.1615(15) & 0.2658(22)(19) \\
7 &  4 & 0.69630 & 0.1519(15) & 0.2745(22)(14) \\
7 &  4 & 0.92840 & 0.14628(74) & 0.2861(20)(03) \\
7 &  4 & 1.23787 & 0.1746(40) & 0.3044(53)(17) \\
7 &  4 & 1.59155 & 0.2486(11) & 0.3284(29)(05) \\
7 &  4 & 1.85681 & 0.3023(23) & 0.3413(39)(34) \\
7 &  4 & 2.22817 & 0.3382(35) & 0.4164(27)(05) \\
7 &  4 & 2.47574 & 0.3603(59) & 0.4425(17)(03) \\
7 &  8 & 2.47574 & 0.1727(38) & 0.2726(40)(67) \\
7 &  4 & 2.78521 & 0.3989(37) & 0.4821(22)(56) \\
7 &  8 & 2.78521 & 0.2096(31) & 0.3272(34)(76) \\
7 &  4 & 3.18310 & 0.4675(19) & 0.5470(56)(13) \\
7 &  8 & 4.45634 & 0.3623(45) & 0.3711(70)(24) \\
7 &  8 & 5.57042 & 0.4359(28) & 0.4596(51)(02) \\
7 & 16 & 5.57042 & 0.2154(30) & 0.2338(24)(21) \\
7 &  8 & 6.36620 & 0.4527(67) & 0.520(03)(18) \\
7 & 16 & 11.14085 & 0.3832(59) & 0.4553(43)(05) \\
\hline
8 &  4 & 0.19894 & 0.1792(10) & 0.2625(26)(01) \\
8 &  4 & 0.55704 & 0.1588(34) & 0.2711(14)(08) \\
8 &  4 & 0.92840 & 0.1476(12) & 0.2993(26)(03) \\
8 &  4 & 1.23787 & 0.1800(33) & 0.3178(23)(03) \\
8 &  4 & 1.59155 & 0.2497(28) & 0.3168(21)(24) \\
8 &  4 & 1.85681 & 0.3007(40) & 0.3144(22)(02) \\
8 &  4 & 2.22817 & 0.2882(25) & 0.3681(29)(08) \\
8 &  4 & 2.47574 & 0.2777(50) & 0.4085(36)(03) \\
8 &  4 & 2.78521 & 0.2567(20) & 0.4683(54)(08) \\
8 &  8 & 2.78521 & 0.1353(24) & 0.2299(30)(27) \\
8 &  4 & 3.18310 & 0.2364(15) & 0.496(13)(27) \\
8 &  8 & 4.45634 & 0.1190(10) & 0.3381(39)(06) \\
8 &  8 & 5.57042 & 0.1474(21) & 0.438(34)(13) \\
8 &  8 & 6.36620 & 0.1871(13) & 0.5191(27)(20) \\
\hline

\\
\end{tabular}
\caption{Mass gap and next excited state energy (in lattice units) in the zero-electric flux scalar sector for $SU(17)$ (continued).}
\label{tab:n17c_glue}
\end{center}
\end{table}

\FloatBarrier

\bibliography{glueballs,spectrum}

\end{document}